\documentclass[11pt]{article}
\pdfoutput=1
\usepackage{epsfig,sint,cite,amsfonts,amssymb,amsmath,bbold}
\usepackage[font=small,labelfont={sf,bf}]{caption}
\usepackage{subcaption}
\usepackage[T1]{fontenc}
\usepackage{tikz}
\usetikzlibrary{decorations.pathreplacing}
\usetikzlibrary{patterns} 

\usepackage[english]{babel}
\usepackage{simplewick}
\usepackage{dsfont}
\usepackage[utf8]{inputenc}

\newcommand{\be}{\begin{equation}}
\newcommand{\ee}{\end{equation}}
\newcommand{\bea}{\begin{eqnarray}}
\newcommand{\eea}{\end{eqnarray}}
\newcommand{\id}{\mathbb{1}}

\newcommand{\tr}{\,\hbox{\rm tr}}

\renewcommand{\vec}[1]{\boldsymbol{#1}}

\begin{document}

\begin{titlepage}
\begin{flushright}
\hfill DESY-16-007\\
\end{flushright}

\vskip 0.5cm

\begin{center}

  {\Large\bf Domain decomposition, multi-level integration and exponential
             noise reduction in lattice QCD\\[0.5ex]} 

\end{center}
\vskip 0.25 cm
\begin{center}

{\large  Marco C\`e}
\vskip 0.125cm
Scuola Normale Superiore, Piazza dei Cavalieri 7, 56126 Pisa, Italy\\
and INFN, Sezione di Pisa, Largo B. Pontecorvo 3, 56127 Pisa, Italy\\
E-mail: marco.ce@sns.it\\
\vskip 0.5cm

{\large  Leonardo Giusti}
\vskip 0.125cm
Dipartimento di Fisica, Universit\`a di Milano--Bicocca,\\
and INFN, sezione di Milano--Bicocca,\\
I-20126 Milano, Italy\\
E-mail: Leonardo.Giusti@mib.infn.it\\
\vskip 0.5cm

{\large  Stefan Schaefer}
\vskip 0.125cm
John von Neumann Institute for Computing (NIC),\\
DESY, Platanenallee 6, D-15738 Zeuthen, Germany\\
E-mail: Stefan.Schaefer@desy.de\\
\vskip 0.5cm

{\bf Abstract}
\vskip 0.35ex
\end{center}

\noindent
We explore the possibility of computing fermionic correlators on
the lattice by combining a domain decomposition with a multi-level
integration scheme. The quark propagator is expanded in series of
terms with a well defined hierarchical structure. The higher the order
of a term, the (exponentially) smaller its magnitude, the less local is
its dependence on the gauge field. Once inserted in a Wick contraction,
the gauge-field dependence of the terms in the resulting series can be
factorized so that it is suitable for multi-level Monte Carlo integration. We
test the strategy in quenched QCD by computing the disconnected correlator
of two flavor-diagonal pseudoscalar densities, and a nucleon two-point function.
In either cases we observe a significant exponential increase of the
signal-to-noise ratio.

\vfill

\eject

\end{titlepage}

\section{Introduction}
With state of the art techniques, the numerical computation of hadronic
correlation functions in lattice Quantum Chromodynamics (QCD) suffers from
signal-to-noise ratios which decrease exponentially with the time separation of
the sources, notable exceptions being the propagators of non-singlet
pseudoscalar mesons. For connected Wick contractions, the problem can be traced
back to the fact that, on a typical gauge configuration, the quark propagator
decreases approximatively as $\exp{\{-M_\pi |y-x|/2\}}$ at asymptotically large distances
$|y-x|$, while the expectation value of a generic hadron correlator decays much
faster~\cite{Parisi:1983ae,Lepage:1989hd}. For a nucleon two-point function at
zero momentum, for instance, the signal-to-noise ratio decreases proportionally
to $\exp{\{-(M_N- 3 M_\pi/2) |y_0-x_0|\}}$, where $|y_0-x_0|$ is the
time-distance of the sources and $(M_N- 3 M_\pi/2)$ is as big as
$3.7$~fm$^{-1}$ at the physical point. The number of configurations needed to
reach a given statistical precision thus increases with the square of that
exponential factor. For disconnected contractions, the problem is even worse
due to vacuum contributions to the variance.

Analogous severe problems afflict the computation of correlators in a large variety
of quantum systems, from the harmonic oscillator to Yang--Mills theory. In some cases,
multi-level algorithms have been proposed which lead to an impressive acceleration of the
simulations~\cite{Albanese:1987ds,Luscher:2001up,Meyer:2002cd,DellaMorte:2007zz,
DellaMorte:2008jd,DellaMorte:2010yp}. They take advantage of the fact that, when the action
and the observables depend locally on the integration variables, the signal-to-noise
problem can be solved by independent measurements of the local building blocks of the
observables. So far, these ideas have been restricted to bosonic theories.

It is not straight forward to formulate multi-level algorithms for systems with
fermions. Once they have been analytically integrated out in the path
integral, the manifest locality of the action and of the observables is lost.
The fermion determinant and propagator are non-local functionals of the background
gauge field. 
In order to make lattice computations with fermions amenable for multi-level
algorithms, factorizations of fermionic correlation functions have to be found,
where the individual terms depend only on gauge links confined to certain
lattice domains. This cannot be achieved with the exact inverse of the Dirac
operator, because each of its elements depends on the gauge
field over the all lattice. As we will see, however, a series of approximations
can be found which exhibit the various degrees of non-locality in the propagator.

In this paper we will pursue two strategies, adapted to different types of
correlation functions, both based on domain decomposition techniques~\cite{Saad:2003,Luscher:2003qa}.
In the first, which we will show to lead to an efficient algorithm for disconnected correlation
functions, we define a succession of domains $\Gamma_0\subset\Gamma_1\subset\Gamma_2$, $\dots$
which are larger and larger and which contain the two end points $x$ and $y$ of the propagator.
The latter can then be expanded in series as 
\be\label{eq:Dseries}
S(y,x) =
S^{(0)}(y,x) + \Big[S^{(1)}-S^{(0)}\Big](y,x) + \Big[S^{(2)}-S^{(1)}\Big](y,x)
+ \ldots\; ,
\ee 
where $S^{(i)}(y,x)$ is the inverse of the Dirac operator restricted to the $i^{\rm th}$ domain,
and depends on the values of the gauge field in $\Gamma_i$ only. The larger the domain $\Gamma_{i}$,
the smaller the corresponding term $[S^{(i+1)}-S^{(i)}]$, the less local is its dependence on
the gauge field. By inserting $S^{(0)}$ in the Wick contraction of the disconnected contribution of
two pseudoscalar densities located in different domains, it is clear that the gauge-field dependence
in the product of the two traces is factorized. The average can then be computed by a two-level Monte
Carlo integration scheme.

The second approach leads one step further, and demonstrates that also connected
hadron correlation functions can be factorized such that multi-level algorithms
can be used. The propagator is approximated by a product of matrices which depend
on the gauge field belonging to different domains of the lattice. This in turn leads to
factorized correlations and thus to local averaging of them.

In the following sections we present the details of the proposed 
computational strategy for the correlation function of two different
flavor-diagonal pseudoscalar densities, and for the nucleon two-point
correlation function. We then show numerical evidence of the effectiveness of the
strategy in quenched QCD, where only the Wick contractions need to be re-organized
in a factorized form. An analogous factorization of the fermion determinant is
left for a future publication.

\section{Quark propagator and locality\label{sec:locality}}
Let us take a lattice $\Gamma$ with open boundary conditions in the
time direction~\cite{Luscher:2011kk}, and define the domains
\be
\Gamma_{0}\subset\Gamma_{1}\subset\Gamma_{2}\subset\dots\subset\Gamma\; . 
\ee
We adopt here the same block 
terminology as in  Ref.~\cite{Luscher:2003qa}.
We choose the $\Gamma_{i}$ to be a hyper-cubic domain of 
lattice points. Its exterior boundary $\partial\Gamma_i$ 
is defined to be the set of all points that have distance 
1 from $\Gamma_{i}$. Each exterior boundary point has a 
closest ``partner'' point in the block. The interior boundary 
$\partial\Gamma^*_i$ of $\Gamma_i$ consists of 
all these points. The set of points that are not in the block is 
denoted by  
\be
\Gamma^*_i =\Gamma\setminus \Gamma_i\; . 
\ee
For a given domain decomposition of the lattice, the Wilson--Dirac 
operator defined in Appendix~\ref{app:Dw}, being a sparse matrix in position space, assumes the 
block-diagonal form 
\be\label{eq:blkdec}
D = 
\left(\begin{array}{c@{~~}c@{~~}}
D_{\Gamma_i} & D_{\partial\Gamma_i}\\
D_{\partial\Gamma^*_i} & D_{\Gamma^*_i}
\end{array}   \right)\; .
\ee
The operator $D_{\Gamma_i}$ acts on quark fields on $\Gamma_i$ in 
the same way as $D$, except that all terms involving the exterior 
boundary points $\partial \Gamma_i$ are set to zero (which is equivalent 
to impose Dirichlet boundary conditions on $\partial\Gamma_i$).
By using the decomposition in Eq.~(\ref{eq:Scmpt}), the exact quark
propagator between the points $x,y\in\Gamma_i$ is given by 
\be\label{eq:bella}
S(y,x) = S^{(i)}(y,x) +\!\!\!\!\! \sum_{w_1,w_2\in\partial\Gamma^*_i}
\!\!\! S^{(i)}(y,w_1)\, 
[D_{\partial\Gamma_i} D^{-1} D_{\partial\Gamma^*_i}](w_1,w_2)\, S^{(i)}(w_2,x)\; ,
\ee
where 
\be\label{eq:diric}
S^{(i)}(y,x) = D^{-1}_{\Gamma_i}(y,x)
\ee
{\it depends on the values of the gauge field in the block $\Gamma_i$ only}.
It is rather clear at this point that we can generate a succession of 
approximations $S^{(i)}$ which, by construction, converges to the exact
propagator when $\Gamma_i$ gets larger and larger. For a typical gauge configuration,
when the sink and the source of the two $S^{(i)}$ in the sum on the r.h.s.\ of Eq.~(\ref{eq:bella})
are at asymptotically large distances, it holds
\be\label{eq:expfaof}
\tr\{S^{(i)}(y,x)\, S^{(i)}(y,x)^\dagger\}^{1/2} \sim
\tr\{S(y,x)\,S(y,x)^\dagger\}^{1/2}\sim e^{-\frac{1}{2} M_\pi |y-x|}
\ee
with $M_\pi$ the mass of the corresponding pseudoscalar meson made of 
degenerate quarks. 
A rough estimate of the distance between the exact and the approximated propagator is 
\be\label{eq:error}
\tr\{(S(y,x) - S^{(i)}(y,x))\,
     (S(y,x)^\dagger - S^{(i)}(y,x)^\dagger) \}^{1/2} \sim e^{- M_\pi d_i}\; ,
\ee
with $d_i$ the average of the distances of $x$ and $y$ from the
interior boundaries of $\Gamma_i$.

\section{Multi-level integration of the disconnected pseudoscalar propagator\label{sec:etap}}
The decomposition in Eq.~(\ref{eq:bella}) calls for
a multi-level integration of disconnected contributions to correlation functions.
We test the idea in SU($3$) Yang--Mills theory with open boundary
conditions~\cite{Luscher:2011kk} supplemented with a doublet of quenched quarks, $u$ and $d$,
degenerate in mass. Both fermions are discretized with the Wilson--Dirac operator,
so that isospin symmetry is exactly preserved. We compute the correlator of two
different flavor-diagonal pseudoscalar densities (the generalization to other cases being straightforward)
\be\label{eq:sing}
C_{P_d}(y_0,x_0) = \frac{1}{L^3}\, \sum_{{\vec x}, {\vec y}}\,
\langle   \bar d(y)\gamma_5 d(y)\,
\bar u(x)\gamma_5 u(x) \rangle =
\frac{1}{L^3}\, \sum_{{\vec x},\, {\vec y}}\, \langle W_{P_d}(y,x) \rangle \; , 
\ee
where $W_{P_d}(y,x)$ indicates the Wick contraction of the fermion fields, and $L$ is the length of the lattice
in the spatial directions\footnote{Throughout the paper dimensionful quantities are always expressed in units
of the lattice spacing $a$ unless explicitly specified.}. In a standard Monte Carlo simulation,
the statistical error of $C_{P_d}(y_0,x_0)$ is constant as a function of $|y_0-x_0|$ while its expectation
value decreases proportionally to $\exp{\{-M_\pi|y_0-x_0|\}}$ at large time separations.
The number of configurations $n_0$ required to reach a given relative statistical error
thus grows exponentially with the time distance of the densities, i.e.\
$n_0 \propto \exp{\{2 M_\pi|y_0-x_0|\}}$.
\begin{figure}[!t]
  \centering
  \begin{subfigure}[t]{0.3\textwidth}
    \centering
    \begin{tikzpicture}
      \draw[dashed] (0,-1) -- (0,+1);
      \begin{scope}[thick, yshift=0.2cm]
        \draw[red,double] (-1.1,0) circle [radius=0.5];
        \fill[white] (-1.6,0) circle [radius=0.12];
        \fill (-1.6,0) circle [radius=0.08] node [left] {$x_0$};
      \end{scope}
      \begin{scope}[thick]
        \draw[red,double] (+1.1,0) circle [radius=0.5];
        \fill[white] (+1.6,0) circle [radius=0.12];
        \fill (+1.6,0) circle [radius=0.08] node [right] {$y_0$};
      \end{scope}
    \end{tikzpicture}
  \end{subfigure}
  \begin{subfigure}[t]{0.3\textwidth}
    \centering
    \begin{tikzpicture}
      \draw[dashed] (0,-1) -- (0,+1);
      \begin{scope}[thick, yshift=0.2cm]
        \draw[red,double] (0,+0.5) -- +(-1.1,0) arc [start angle=90, end angle=180, radius=0.5] ;
        \draw[red,double] (0,-0.5) -- +(-1.1,0) arc [start angle=270, end angle=180, radius=0.5] ; 
        \draw (0,-0.5) arc [start angle=-90, end angle=90, radius=0.5];
        \fill[white] (-1.6,0) circle [radius=0.12];
        \fill (-1.6,0) circle (0.08) node [left] {$x_0$};
        \fill[white] (0,+0.5) circle [radius=0.08];
        \fill (0,+0.5) circle (0.04);
        \fill[white] (0,-0.5) circle [radius=0.08];
        \fill (0,-0.5) circle (0.04);
      \end{scope}
      \begin{scope}[thick]
        \draw[red,double] (+1.1,0) circle [radius=0.5];
        \fill[white] (+1.6,0) circle [radius=0.12];
        \fill (+1.6,0) circle [radius=0.08] node [right] {$y_0$};
      \end{scope}
    \end{tikzpicture}
  \end{subfigure}
  \begin{subfigure}[t]{0.3\textwidth}
    \centering
    \begin{tikzpicture}
      \draw[dashed] (0,-1) -- (0,+1);
      \begin{scope}[thick, yshift=0.2cm]
        \draw[red,double] (0,+0.5) -- +(-1.1,0) arc [start angle=90, end angle=180, radius=0.5] ;
        \draw[red,double] (0,-0.5) -- +(-1.1,0) arc [start angle=270, end angle=180, radius=0.5] ; 
        \draw (0,-0.5) arc [start angle=-90, end angle=90, radius=0.5];
        \fill[white] (-1.6,0) circle [radius=0.12];
        \fill (-1.6,0) circle (0.08) node [left] {$x_0$};
        \fill[white] (0,+0.5) circle [radius=0.08];
        \fill (0,+0.5) circle (0.04);
        \fill[white] (0,-0.5) circle [radius=0.08];
        \fill (0,-0.5) circle (0.04);
      \end{scope}
      \begin{scope}[thick]
        \draw[red,double] (0,+0.5) -- (+1.1,+0.5) arc [start angle=90, end angle=0, radius=0.5] ;        
        \draw[red,double] (0,-0.5) -- (+1.1,-0.5) arc [start angle=-90, end angle=0, radius=0.5] ;
        \draw   (0,+0.5) arc [start angle=90, end angle=270, radius=0.5];
        \fill[white] (+1.6,0) circle [radius=0.12];
        \fill (+1.6,0) circle [radius=0.08] node [right] {$y_0$};
        \fill[white] (0,+0.5) circle [radius=0.08];
        \fill (0,+0.5) circle (0.04);
        \fill[white] (0,-0.5) circle [radius=0.08];
        \fill (0,-0.5) circle (0.04);
      \end{scope}
    \end{tikzpicture}
  \end{subfigure}
  \caption{The three type of contributions to the disconnected pseudoscalar propagator
in Eqs.~(\protect\ref{eq:W0})--(\protect\ref{eq:rall}).
           Black (single) lines are full propagators, red (double) ones are those within a domain.}
  \label{eq:figsing}
\end{figure}
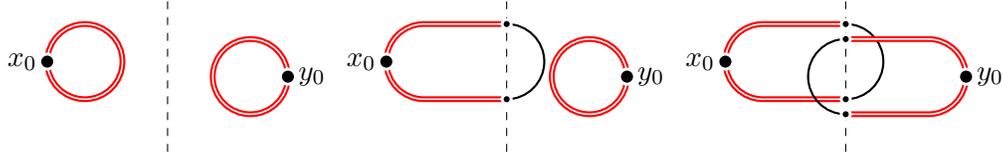

The depletion of the signal-to-noise ratio is particularly severe at large time-distances.
To take advantage of the locality of the theory, it is therefore natural to divide the
lattice $\Gamma$ in two non-overlapping thick time-slices $\Gamma_0$ and $\Gamma^*_0$.
The first time coordinate of $\Gamma^*_0$, $x_0^{\rm cut}$, is chosen approximatively in the
middle between the two densities (see Fig.~\ref{eq:figsing}).
By using the first and the second diagonal elements in Eqs.~(\ref{eq:Scmpt}) and
Eqs.~(\ref{eq:Scmpt2}) respectively, the Wick contraction can be decomposed as
\be\label{eq:deco1}
W_{P_d}(y,x) = W^{(\text{f})}_{P_d}(y,x) + W^{(\text{r})}_{P_d}(y,x)\; , 
\ee
where 
\be\label{eq:W0}
W^{(\text{f})}_{P_d}(y,x) = \tr\Big\{\gamma_5 D^{-1}_{\Gamma_0}(x,x)\Big\} \times
\tr\Big\{\gamma_5 D^{-1}_{\Gamma^*_0}(y,y)\Big\}\; ,
\ee
$x\in \Gamma_0$, and $y\in\Gamma^*_0$. The rest of the contraction
is given by
\be
W^{(\text{r})}_{P_d}(y,x) = \Big[W^{(\text{r}_1)}_{P_d}(y,x) + (\Gamma_0,x) \leftrightarrow (\Gamma^*_0,y) \Big]
+ W^{(\text{r}_2)}_{P_d}(y,x)
\ee
with
\bea
\!\!\!\!\!\!\!\!\!\!W^{(\text{r}_1)}_{P_d}(y,x)\!\!\! & = &\!\!\!\tr\Big\{\gamma_5 D^{-1}_{\Gamma_0}(x,\cdot)
[D_{\partial\Gamma_0} D^{-1} D_{\partial\Gamma^*_0}](\cdot,\cdot)D^{-1}_{\Gamma_0}(\cdot,x)\Big\}
\times
\tr\Big\{\gamma_5 D^{-1}_{\Gamma^*_0}(y,y)\Big\},\label{eq:rall0}\\[0.5cm]
\!\!\!\!\!\!\!\!\!\!W^{(\text{r}_2)}_{P_d}(y,x)\!\!\! & = &\!\!\!\tr\Big\{\gamma_5 D^{-1}_{\Gamma_0}(x,\cdot)
[D_{\partial\Gamma_0} D^{-1} D_{\partial\Gamma^*_0}](\cdot,\cdot)D^{-1}_{\Gamma_0}(\cdot,x)\Big\}
\times\nonumber\\[0.25cm]
& &\!\!\! \tr\Big\{\gamma_5 D^{-1}_{\Gamma^*_0}(y,\cdot) [D_{\partial\Gamma^*_0} D^{-1}
D_{\partial\Gamma_0}](\cdot,\cdot)D^{-1}_{\Gamma^*_0}(\cdot,y)\Big\}\; .\label{eq:rall}
\eea
When the spatial gauge links at $x_0^{\rm cut}$ are kept frozen, 
the dependence of the action and of $W^{(\text{f})}_{P_d}(y,x)$ on the remaining link variables
is factorized.

\subsection{Two-level integration}
When an observable depends only on the link variables in a given sub-lattice
and the action of the theory is local, it is useful to define its expectation
value restricted to that domain. This is a function of the link variables at the
boundary of the sub-lattice only, and do not depend on the gauge field values elsewhere. 
For the trace of the Wilson--Dirac operator that we are interested in, it reads
\be
\left[\tr\left\{ \gamma_5 D^{-1}_{\Gamma_0}(x,x) \right\} \right] =
\frac{1}{Z_{\Gamma_0}} \int D[U]_{\Gamma_0} e^{-S[U]_{\Gamma_0}} \tr\left\{ \gamma_5 D^{-1}_{\Gamma_0}(x,x) \right\}
\ee
where $D[U]_{\Gamma_0}$ and $S[U]_{\Gamma_0}$ are the invariant Haar measure and the action
restricted
to the domain $\Gamma_0$, and the sub-lattice partition function is fixed by requiring that 
$[1]=1$. By following the standard line of argumentation in multi-level integration
technique~\cite{Albanese:1987ds,Luscher:2001up,Meyer:2002cd,DellaMorte:2007zz,DellaMorte:2008jd,DellaMorte:2010yp},
it follows that
\be
\left\langle \!\!\tr\left\{\! \gamma_5 D^{-1}_{\Gamma_0}(x,x) \right\} 
             \!\!\tr\left\{\! \gamma_5 D^{-1}_{\Gamma^*_0}(y,y)   \right\}
\right\rangle\! = 
\left\langle \left[\!\tr\left\{\! \gamma_5 D^{-1}_{\Gamma_0}(x,x) \right\} \right]\!\!
                 \left[\!\tr\left\{\! \gamma_5 D^{-1}_{\Gamma^*_0}(y,y)   \right\} \right] \right\rangle .
\ee
This suggests that the mean value of $W^{(\text{f})}_{P_d}(y,x)$ can be computed with a
two-level algorithm. First, for each level-0 gauge field (spatial components) on the
boundary $x_0^{\rm cut}$, the averages
of the two traces are carried out independently on $n_1$ level-1 configurations
generated independently in the two thick time-slices. Then the average over the level-0
configurations of the product of the two means is performed by updating the gauge links
over the entire lattice. The crucial question, to be answered numerically in section~\ref{sec:etapN},
is whether one can choose $n_1$ large enough to profit from the level-1 averaging, or
if instead the variance of the factorized contribution is dominated by the fluctuations
of the spatial components of the gauge field at the boundary. If $n_1$ can be taken large
enough such
that the product of the (level-1) mean values is proportional to $\exp{\{-M_\pi|y_0-x_0|\}}$,
then a good statistical precision is reached with a number of updates 
$(n_{0} \cdot n_{1})\propto \exp{\{M_\pi|y_0-x_0|\}}$. Notice that the factor in the exponent
is halved with respect to the standard Monte Carlo.\\

The contribution from $W^{(\text{r}_1)}_{P_d}(y,x)$ is expected to be
suppressed, for a typical configuration,  by a factor
$\exp\{-M_\pi |x_0-x_0^{\rm cut}|\}$ at large time separations. Measuring
it over the $n_0 \cdot n_1$ configurations generated in the two-level update,
by blocking the results and averaging over the $n_0$ of them, may be enough to reduce the
error at the same level of the one of $W^{(\text{f})}_{P_d}(y,x)$ (up to a pre-factor that have to
be quantified numerically). The last contribution, $W^{(\text{r}_2)}_{P_d}(y,x)$, is expected to
be already proportional to $\exp\{-M_\pi |y_0-x_0|\}$. This is of the same order of the
expected signal, and therefore the standard level-0 average is adequate.

\section{Factorization of the approximated quark propagator\label{sec:fact}}
\begin{figure}[!t]
  \centering
  \begin{tikzpicture}
    \begin{scope}
      \clip (-0.3,-1.8) rectangle (13.8,+1.8);
      \draw[step=0.5, dotted] (-4,-2.5) grid (18,+2.5);
      \foreach \x in {-4,-2,...,18}
        \draw[dashed] (\x-0.25,-2.5) -- (\x-0.25,+2.5);
    \end{scope}
    \begin{scope}[thick, xshift=2.75cm]
      \draw (0,0) .. controls (1,1) and (7,1) .. (8,0);
      \fill[white] (0,0) circle [radius=0.12];
      \fill (0,0) circle [radius=0.08] node [left] {$x_0$};
      \fill[white] (8,0) circle [radius=0.12];
      \fill (8,0) circle [radius=0.08] node [right] {$y_0$};
    \end{scope}
    \draw[decorate, decoration={brace, amplitude=5}] (-0.25,1.85) -- +(12,0) node [midway, above=3] {$\Omega_{l+1}$};
    \begin{scope}[xshift=1cm, yshift=-1.5cm]
      \draw ( 0,0) node {$\Lambda_{m-1}$};
      \draw ( 2,0) node {$\Lambda_m$};
      \draw ( 4,0) node {$\Lambda_{m+1}$};
      \draw ( 6,0) node {$\Lambda_{\dots}$};
      \draw ( 8,0) node {$\Lambda_{l-1}$};
      \draw (10,0) node {$\Lambda_l$};
      \draw (12,0) node {$\Lambda_{l+1}$};
    \end{scope}
    \begin{scope}[xshift=-0.25cm, yshift=-1.85cm]
      \draw[decorate, decoration={brace, amplitude=5, mirror}] (0,0) -- +(4,0) node [midway, below=3] {$\Omega^*_{m-1}$};
      \draw[decorate, decoration={brace, amplitude=5, mirror}] (4,0) -- +(4,0) node [midway, below=3] {$\Omega^*_{m+1}$};
      \draw[decorate, decoration={brace, amplitude=5, mirror}] (8,0) -- +(4,0) node [midway, below=3] {$\Omega^*_{l-1}$};
    \end{scope}
    \begin{scope}[xshift=1.75cm, yshift=-2.5cm]
      \draw[decorate, decoration={brace, amplitude=5, mirror}] (0,0) -- +(4,0) node [midway, below=3] {$\Omega^*_m$};
      \draw[decorate, decoration={brace, amplitude=5, mirror}] (4,0) -- +(4,0) node [midway, below=3] {$\Omega^*_{\dots}$};
      \draw[decorate, decoration={brace, amplitude=5, mirror}] (8,0) -- +(4,0) node [midway, below=3] {$\Omega^*_l$};
    \end{scope}
  \end{tikzpicture}
  \caption{Domain decomposition of the lattice in thick time-slices, with the sink and the
    source of the quark propagator belonging to blocks distant in time.}
  \label{Fig:fig3}
\end{figure}
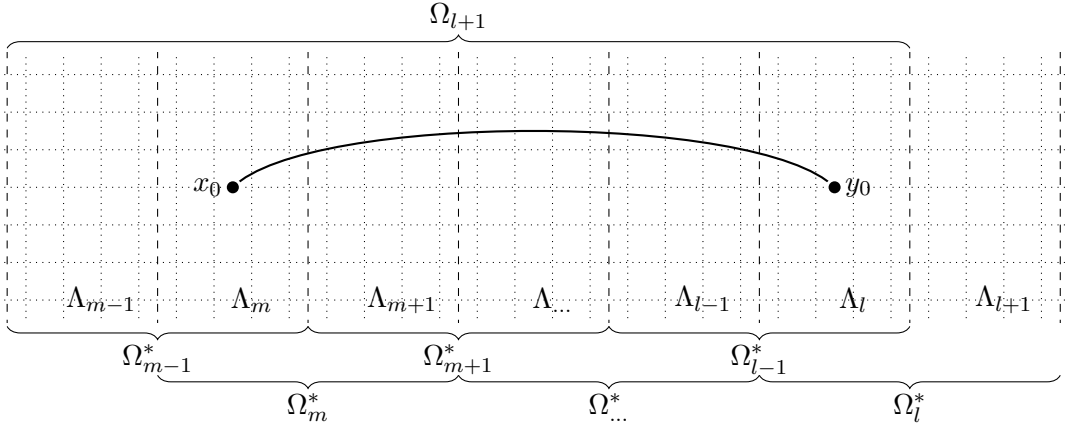

The decomposition discussed in section~\ref{sec:locality} can be generalized
for approximating the propagator between two points with a large temporal
separation. A simple domain decomposition, where this can be done in
practice, is the one where the lattice is divided in thick time-slices
$\Lambda_i$ all\footnote{We choose this set up for simplicity. The lattice
can, of course, be divided in domains of different sizes if required by a
specific problem.}
of thickness $\Delta$, with
$i=0,\dots,n-1$, $n=T/\Delta$ and $T$ being the time-extension
of the lattice (see Fig.~\ref{Fig:fig3}). The block
structure of the Wilson--Dirac matrix\footnote{The same decomposition applies
to the $O(a)$-improved Wilson-Dirac operator as well.} is then given by 
\be\label{eq:fact1}
D = \left(\begin{array}{c@{~~}c@{~~}c@{~~}c@{~~}c@{~~}c}
\dots  & \dots & D_{\Lambda_{i-1,i}} & 0 & 0 & 0\\
0 & D_{\Lambda_{i,i-1}} & D_{\Lambda_{i,i}} & D_{\Lambda_{i,i+1}} & 0 & 0\\
0 & 0 & D_{\Lambda_{i+1,i}} & D_{\Lambda_{i+1,i+1}} & D_{\Lambda_{i+1,i+2}}& 0\\
0 & 0 & 0 & D_{\Lambda_{i+2,i+1}}  & \dots & \dots
\end{array}   \right)\; ,
\ee
where $D_{\Lambda_{i,i}}$ acts on quark fields in $\Lambda_{i}$ in the same way as $D$, except that all terms
involving the exterior boundary points are set to zero. The off-diagonal terms on the r.h.s.\ of
Eq.~(\ref{eq:fact1}) are given by
\bea\label{eq:fact3}
D_{\Lambda_{i,i-1}}(\vec x,\vec y) & = &  P_+ U_0^\dagger(x_0^i-1,\vec x)\, \delta_{\vec x,\vec y}
\; , \\[0.25cm]
D_{\Lambda_{i,i+1}}(\vec x,\vec y) & = & P_- U_0(x_0^{i+1}-1,\vec x)\, 
\delta_{\vec x,\vec y}\; ,\nonumber 
\eea
where $U_0(x)$ are the temporal links, $P_{\pm} = (1\pm\gamma_0)/2$, and
$x_0^j=j\cdot\Delta$ is the first time slice of the block $\Lambda_i$. By
using Eq.~(\ref{eq:g5h}), it is easy to show that 
\be\label{eq:g5Hermite}
D_{\Lambda_{i,i}} = \gamma_5\, D^\dagger_{\Lambda_{i,i}}\, \gamma_5\;, \qquad 
D_{\Lambda_{i,i-1}} = 
\gamma_5\, D^\dagger_{\Lambda_{i-1,i}}\, \gamma_5\;.
\ee
The blocking in Eq.~(\ref{eq:fact1}) and the decomposition in Eq.~(\ref{eq:Scmpt2}),
or equivalently in Eq.~(\ref{eq:Scmpt}), are the basic ingredients for constructing
an approximated propagator between two points whose distance is much larger
than $\Delta$. This can be achieved as described in the following three steps.\\

\noindent {\bf Step 1:} If $x\in\Lambda_m$ and $y\in\Lambda_l$, with $l>m$ for instance, choose 
\be
\Gamma_0 =\Lambda_{m-1}\cup \Lambda_m \cup\dots\cup
\Lambda_{l}\cup\Lambda_{l+1}\; .
\ee
Thanks to the results in section \ref{sec:locality}
\be\label{eq:stp1}
S(y,x) = D^{-1}_{\Gamma_0}(y,x)+\dots\; .
\ee

\noindent {\bf Step 2:} Decompose $\Gamma_0$ in overlapping domains
\be
\Omega_i=\Lambda_{m-1}\cup\Lambda_{m}\cup\dots\cup\Lambda_{i-1}\; , 
\qquad i=m+2,\dots,l+1\; ,
\ee
and\footnote{Notice that  $\Omega_l\cup\Omega^*_l=\Gamma_0$ is valid for $i=l$ only.}  
\be
\Omega^*_i=\Lambda_{i}\cup\Lambda_{i+1}\; ,
\ee
with the boundary operators of $\Omega_i$ given by 
\be
D_{\partial\Omega_i} = D_{\Lambda_{m-1,m-2}} + D_{\Lambda_{i-1,i}}\; ,\qquad
D_{\partial\Omega^*_i} = D_{\Lambda_{m-2,m-1}} + D_{\Lambda_{i,i-1}}\; .
\ee
By taking the bottom-left off-diagonal element in
Eq.~(\ref{eq:Scmpt2}), one arrives at 
\be\label{eq:stp2}
D^{-1}_{\Gamma_0}(y,x) = -\!\!\!\!\! \sum_{\tiny\begin{array}{c}
w_1\in\partial\Omega_l\\[0.075cm]
w_2\in\partial\Omega^*_l\end{array}}\!\!\!\!\!\! 
D^{-1}_{\Omega^*_l}(y,w_1)\, 
D_{\partial\Omega^*_l}(w_1,w_2)\, D^{-1}_{\Gamma_0}(w_2,x)\; . 
\ee

\noindent {\bf Step 3:} 
Since $w_2$ and $x$ in Eq.~(\ref{eq:stp2}) are both at least at a 
distance $\Delta$ from the exterior boundary of $\Omega_{l+1}$,
one can replace 
\be
D^{-1}_{\Gamma_0}(w_2,x)= D^{-1}_{\Omega_{l+1}}(w_2,x)+\dots\; 
\ee
and arrive to 
\be
D^{-1}_{\Gamma_0}(y,x) = -\!\!\!\!\! \sum_{\tiny\begin{array}{c}
w_1\in\partial\Omega_l\\[0.075cm]
w_2\in\partial\Omega^*_l\end{array}}\!\!\!\!\!\! D^{-1}_{\Omega^*_l}(y,w_1)\, 
D_{\Lambda_{l,l-1}}(w_1,w_2)\, D^{-1}_{\Omega_{l+1}}(w_2,x) +\dots\; .  
\label{dec1}
\ee
The boundary operator $D_{\partial\Omega^*_l}$ has been replaced by 
$D_{\Lambda_{l,l-1}}$ since this is the only component 
acting on fields in $\Omega^*_{l}$.\\

\noindent By iterating $(m-l)$ times steps 2 and 3, it is easy to show that
one can define an approximated propagator
\be\label{eq:fctpr}
S^{(\text{f})}(y,x) = (-1)^{m-l} \Big[\prod^{m+1}_{i=l} D^{-1}_{\Omega^*_i}\, 
D_{\Lambda_{i,i-1}}\Big](y,\cdot)\, D^{-1}_{\Omega_{m+2}}(\cdot,x) 
\ee
which satisfies $\gamma_5$-hermiticity. Since in each
step the (inverse) matrix factors have been approximated so that the source
and the sink coordinates are at least at a distance $\Delta$ from the
Dirichlet boundary conditions, we expect 
\be\label{eq:errorfact}
\tr\{(S(y,x) - S^{(\text{f})}(y,x))\,
     (S(y,x)^\dagger - S^{(\text{f})}(y,x)^\dagger) \}^{1/2} \sim e^{-M_\pi \Delta}\; .
\ee
The crucial property of the r.h.s.\ of Eq.~(\ref{eq:fctpr}) is that the dependence 
on the gauge field is factorized. The various propagators $D^{-1}_{\Omega^*_i}$ 
depend on the values taken by the gauge field in two thick slices only, while 
the last one $D^{-1}_{\Omega_{m+2}}$ on three. Remarkably the formula 
(\ref{eq:fctpr}) is a systematic approximation of the {\it exact} formula 
in Eq.~(\ref{eq:LUf}) derived from the LU decomposition of the Dirac 
operator, see Appendix \ref{eq:appLU}. A succession of approximations
of the type in Eq.~(\ref{eq:Dseries}) can finally be constructed by taking larger and
larger values of $\Delta$.

\subsection{Factorization and baryon symmetry}
To insert baryon projectors in the partition function at 
the boundaries of the blocks, we can introduce in the action
of the theory a time-dependent $U(1)_V$ field $\theta(x_0)$ 
constant in space
\be
\theta(x_0) = \left\{\begin{array}{cc}
\theta_i &  x_0=\Delta\, (i+1)-1\\[0.375cm]
0 & x_0\neq \Delta\, (i+1)-1
\end{array}\right.
\ee 
by replacing 
\be
D_{\Lambda_{i,i+1}}\longrightarrow e^{-i\theta_i} D_{\Lambda_{i,i+1}}\;, \quad
D_{\Lambda_{i+1,i}}\longrightarrow e^{i\theta_i} D_{\Lambda_{i+1,i}}\; .
\ee
Since the $A_i$ in Eq.~(\ref{e:B}) are $\theta$-independent, 
the dependence on $\theta$ of the exact propagator can be easily 
deduced from the Eq.~(\ref{eq:psil}). It is given by  
\be\label{eq:thetatr}
S(y,x) \longrightarrow 
\exp{\Big\{\displaystyle i 
\sum_{i=m}^{l-1} \theta_i\Big\}}\, S(y,x)\;,\;\;\;
x\in\Lambda_m\;,\;\; y\in\Lambda_l\;,\;\; l>m\; . 
\ee
The $D_{\Lambda_{i,i}}$ are also $\theta$-independent, and therefore the approximated
quark propagator $S^{(\text{f})}(y,x)$ in Eq.~(\ref{eq:fctpr}) inherits the very
same $\theta$-dependence of the exact one in Eq.~(\ref{eq:thetatr}), i.e.\ the approximation
preserves the baryon symmetry.

\section{Multi-level integration of pion and baryon propagators}
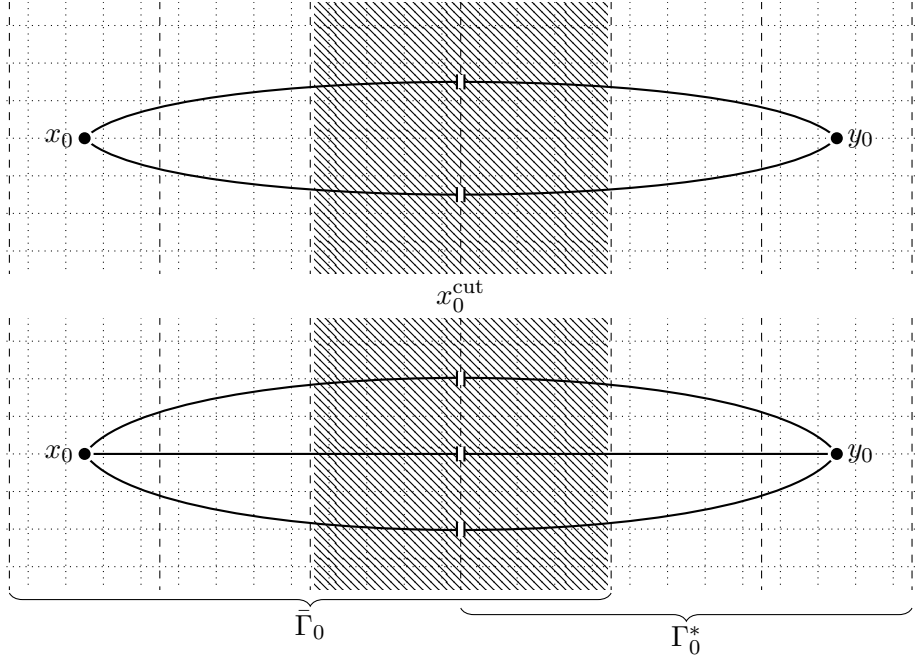
\begin{figure}[t!]
  \centering
  \begin{tikzpicture}
    \begin{scope}[yshift=4.2cm]
      \clip (-0.3,-1.8) rectangle (11.8,+1.8);
      \draw[step=0.5, dotted] (-4,-2.5) grid (18,+2.5);
      \foreach \x in {-4,-2,...,18}
        \draw[dashed] (\x-0.25,-2.5) -- (\x-0.25,+2.5);
      \fill[pattern=north west lines] (3.8,-2.5) rectangle (7.7,+2.5);
    \end{scope}
    \begin{scope}[thick, xshift=0.75cm, yshift=4.2cm]
      \draw (0,0) .. controls (1,+1) and (9,+1) .. (10,0);
      \draw (0,0) .. controls (1,-1) and (9,-1) .. (10,0);
      \fill[white] ( 0,0) circle [radius=0.12];
      \fill ( 0,0) circle [radius=0.08] node [left] {$x_0$};
      \fill[white] (10,0) circle [radius=0.12];
      \fill (10,0) circle [radius=0.08] node [right] {$y_0$};
      \begin{scope}[xshift=5cm]
        \foreach \y in {-0.75,+0.75}
        {
          \fill[white] (-0.05,\y-0.08) rectangle (+0.05,\y+0.08);
          \draw (-0.05,\y-0.1) -- (-0.05,\y+0.1);
          \draw (+0.05,\y-0.1) -- (+0.05,\y+0.1);
        }
      \end{scope}
      \draw (5,-2.1) node {$x_0^\text{cut}$};
    \end{scope}
    \begin{scope}
      \clip (-0.3,-1.8) rectangle (11.8,+1.8);
      \draw[step=0.5, dotted] (-4,-2.5) grid (18,+2.5);
      \foreach \x in {-4,-2,...,18}
        \draw[dashed] (\x-0.25,-2.5) -- (\x-0.25,+2.5);
      \fill[pattern=north west lines] (3.8,-2.5) rectangle (7.7,+2.5);
    \end{scope}
    \begin{scope}[thick, xshift=0.75cm]
      \draw (0,0) .. controls (1,+1.35) and (9,+1.35) .. (10,0);
      \draw (0,0) -- (10,0);
      \draw (0,0) .. controls (1,-1.35) and (9,-1.35) .. (10,0);
      \fill[white] ( 0,0) circle [radius=0.12];
      \fill ( 0,0) circle [radius=0.08] node [left] {$x_0$};
      \fill[white] (10,0) circle [radius=0.12];
      \fill (10,0) circle [radius=0.08] node [right] {$y_0$};
      \begin{scope}[xshift=5cm]
        \foreach \y in {-1,0,+1}
        {
          \fill[white] (-0.05,\y-0.08) rectangle (+0.05,\y+0.08);
          \draw (-0.05,\y-0.1) -- (-0.05,\y+0.1);
          \draw (+0.05,\y-0.1) -- (+0.05,\y+0.1);
        }
      \end{scope}
    \end{scope}
    \draw[decorate, decoration={brace, amplitude=5, mirror}] (-0.25,-1.85) -- +(8,0) node [midway, below=3] {$\bar\Gamma_0$};
    \draw[decorate, decoration={brace, amplitude=5, mirror}] (+5.75,-2.05) -- +(6,0) node [midway, below=3] {$\Gamma^*_0$};
  \end{tikzpicture}
  \caption{Sketch of the factorized approximation of the pion and baryon propagators.}
  \label{fig:4}
\end{figure}

There is no unique way to design a multi-level integration algorithm
by starting from Eqs.~(\ref{eq:stp2})--(\ref{eq:fctpr}). A
simple possibility to start with is to divide the full lattice $\Gamma$ in only two
overlapping thick time-slices: $\bar\Gamma_0$ which includes the time slices
$[0,x_0^{\rm cut}+\Delta-1]$ (it includes one more thick time-slice of thickness
$\Delta$ with respect to the $\Gamma_0$ defined in section~\ref{sec:etap}),
where  $x_0^{\rm cut} = i_c\, \Delta$, and
$\Gamma^*_0$ which include those in $[x_0^{\rm cut}, T-1]$. Open boundary conditions
in time for the full lattice are again assumed for simplicity. By taking the bottom-left
off-diagonal element of the decomposition (\ref{eq:Scmpt2}), and by replacing the
full propagator with $D^{-1}_{\bar\Gamma_0}$, the approximated factorized propagator
can be written as
\be\label{eq:approxlong}
S^{(\text{f})}(y,x) = - D^{-1}_{\Gamma^*_0}(y,\cdot) D_{\partial \Gamma^*_0}(\cdot,\cdot)
D^{-1}_{\bar\Gamma_0}(\cdot,x)
\ee
where $x \in \bar\Gamma_0$ and $y \in\Gamma^*_0$, see Fig.~\ref{fig:4}.
In this setup {\it $\Delta$ is simply the thickness of $\bar\Gamma_0\cap\Gamma^*_0$}.
In order to cut the fermion lines in the Wick contractions, so to transform the matrix
products into ordinary ones, we introduce the projector 
\be\label{eq:proj}
P_L (y,x) = \sum_{i=1}^{N_m}\,
\phi_i(x) \, \phi_i^\dagger(y)
\ee
where $\phi_i$ are $N_m$ orthonormal vectors. The projector is then used to define a further
approximated propagator 
\be\label{eq:approxlong2}
\tilde S^{(\text{f})}(y,x) = - \sum_{i}\, [D^{-1}_{\Gamma^*_0}
D_{\partial \Gamma^*_0}\phi_i](y)\, [\phi_i^\dagger D^{-1}_{\bar\Gamma_0}](x)\; . 
\ee
In the following sections we will use two different set of vectors $\phi_i$: those
which span the deflation subspace as defined in Ref.~\cite{Luscher:2007se},
and $N_m$ orthonormal vectors constructed by applying $10$ inverse iterations
of the Wilson-Dirac operator defined in the domain $\Omega^*_{i_c-1}=\Lambda_{i_c-1}\cup\Lambda_{i_c}$
with Dirichlet boundary conditions on its exterior boundaries.

\subsection{Connected pseudoscalar propagator}
The correlation function of two flavor non-diagonal pseudoscalar densities is defined as 
\be\label{eq:pipi}
C_{P_c}(y_0,x_0) = \frac{1}{L^3}\, \sum_{{\vec x}, {\vec y}}\,
\langle   \bar d(y)\gamma_5 u(y)\,
\bar u(x)\gamma_5 d(x) \rangle = -
\frac{1}{L^3}\, \sum_{{\vec x}, {\vec y}}\, \langle W_{P_c}(y,x) \rangle \; ,
\ee
where $W_{P_c}(y,x)$ is the corresponding Wick contraction. By using the
factorized propagator in Eq.~(\ref{eq:approxlong}), the approximated
Wick contraction can be written as
\be\label{eq:W0pc}
W^{(\text{f})}_{P_c}(y,x) = 
\tr\Big\{S^{(\text{f})}(y,x) S^{(\text{f})\, \dagger}(y,x)\Big\}\, . 
\ee
When $\bar\Gamma_0$, or equivalently $\Delta$, gets larger and larger, the Eq.~(\ref{eq:W0pc})
generates a succession of approximations whose rest
\be\label{eq:W0pc2}
W^{(\text{r})}_{P_c}(y,x) = W_{P_c}(y,x) - W^{(\text{f})}_{P_c}(y,x) 
\ee
converges exponentially fast to zero. Even if the
gauge field appears in a factorized form, it is difficult to implement a
multi-level integration scheme by using Eq.~(\ref{eq:W0pc}). To this aim
we can use the approximated propagator in Eq.~(\ref{eq:approxlong2}), and define
\be\label{eq:fullfactpi}
\tilde W^{(\text{f})}_{P_c}(y,x) = \sum_{i,j} \tilde B_{ij}(y) B_{ji}(x)
\ee
where
\begin{align}
\label{eq:Bpions}
B_{ij}(x) & =  [\phi_i^\dagger D^{-1}_{\bar\Gamma_0}\gamma_5](x)\,
                                 [D^{-1}_{\bar\Gamma_0}\gamma_5\phi_j](x)\; ,\nonumber\\[0.25cm]
\tilde B_{ij}(y) & = 
                      [\phi_i^\dagger \gamma_5 D_{\partial\Gamma^*_0}
                       D^{-1}_{\Gamma^*_0}\gamma_5](y)\,
                        [D^{-1}_{\Gamma^*_0} D_{\partial\Gamma^*_0} \phi_j](y)\; , 
\end{align}
and the vectors $\phi_i$ depend only on the gauge links belonging to the
thick time-slice $\Omega^*_{i_c-1}$.
When the gauge links in $\Omega^*_{i_c-1}$ are kept frozen, 
the dependence of the action and of $\tilde W^{(\text{f})}_{P_c}(y,x)$ on the remaining link variables
in $\bar\Gamma_0$ and $\Gamma^*_0$ is factorized. A multi-level similar to the one designed in
section~\ref{sec:etap} can thus be easily implemented.

\subsection{Nucleon propagator}
A possible choice for baryon interpolating operators is  
\be
N = \Big[u^{a T} C \gamma_5 d^b \Big] d^c \epsilon^{abc}\,, 
\qquad
\bar N = \bar d^e \Big[\bar d^f  C \gamma_5  \bar u^{g T}\Big]  \epsilon^{feg}\,. 
\ee
The corresponding two-point function reads
\be
C_N(y_0,x_0)\! =\!  \frac{1}{L^3} \sum_{{\bf  x, y}}
\tr  \Big[\, \langle N(y) \bar N(x)\rangle P_-\, \Big]\! =\! 
\frac{1}{L^3} \sum_{{\bf x,y}}
\Big\{\langle W_{N1}(y,x) \rangle\! -\! \langle W_{N2}(y,x)\rangle \Big\}\; , 
\ee
where $C=i\gamma_0\gamma_2$, and the trace is over the nucleon spinor 
indices. The $W_{N1}(y,x)$ and $W_{N2}(y,x)$ are the two Wick contractions
\begin{align}
\label{eq:baryfact}
W_{N1}(y,x) & =  \tr[S_u^{ag\, T}(y,x) C \gamma_5 S_d^{bf}(y,x) C \gamma_5] 
\tr[S_d^{ce}(y,x)P_-] \epsilon^{abc} \epsilon^{feg}\;,\nonumber\\[0.375cm]
W_{N2}(x,y) & =  \tr[S_u^{ag\, T}(y,x) C \gamma_5 S_d^{be}(y,x)P_- 
S_d^{cf}(y,x) C \gamma_5] \epsilon^{abc} \epsilon^{feg}\; ,
\end{align}
where again the trace is on the spinor index only. The factorized approximation
of the nucleon Wick contractions $W^{(\text{f})}_{N1}$ and $W^{(\text{f})}_{N2}$ are defined
analogously to $W_{N1}$ and $W_{N2}$ by replacing on the r.h.s.\ of Eqs.~(\ref{eq:baryfact})
each quark propagator by its approximated factorized expression in Eq.~(\ref{eq:approxlong}). Finally 
the corresponding $\tilde W^{(\text{f})}_{N1}$ and $\tilde W^{(\text{f})}_{N2}$ are defined as
\begin{align}
\label{eq:W0tbary}
\tilde W^{(\text{f})}_{N1}(y,x) & =  - \sum_{ijk}\mathds{B}[\xi_j,\xi_i,\xi_k;x]^T \gamma_5 P_-
\mathds{B}[\eta_i,\eta_j,\eta_k;y]\nonumber\\[0.25cm]
\tilde W^{(\text{f})}_{N2}(y,x) & =  - \sum_{ijk}\ \mathds{B}[\xi_k,\xi_i,\xi_j;x]^T \gamma_5 P_-
\mathds{B}[\eta_i,\eta_j,\eta_k;y]
\end{align}
where 
\be
\xi^T_i = \phi_i^\dagger D^{-1}_{\bar\Gamma_0}\; , \qquad
\eta_i= D^{-1}_{\Gamma^*_0}D_{\partial\Gamma^*_0}\phi_i\; , 
\ee
and the colorless spinors 
\be
\mathds{B}[s_i,s_j,s_k;x]_\alpha =  \epsilon^{abc}\Big\{
[s^T_i]^a(x)\, C \gamma_5\, [s_j]^b(x)\Big\} [s_k]^c_\alpha(x)
\ee
have been introduced. By choosing again the vectors $\phi_i$ to depend only on
the gauge links belonging to the thick time-slice $\Omega^*_{i_c-1}$,
the two colorless spinors on the r.h.s.\ of
Eqs.~(\ref{eq:W0tbary}) depend only on the gauge links belonging
to $\bar\Gamma_0$ and $\Gamma^*_0\cup\Omega^*_{i_c-1}$ respectively. When the links
belonging to $\Omega^*_{i_c-1}$ are kept frozen, 
the dependence of the action and of the two approximated Wick
contractions is factorized. Also in this case their mean value
can then be computed with a two-level algorithm analogous to what
has been described in section~\ref{sec:etap} for the disconnected
pseudoscalar propagator. Contrary to the pion propagator, the
signal-to-noise ratio in the baryon correlator is exponentially
suppressed with the distance of the sources. If it turns out to be
profitable to choose $n_1$ so that the product of the (level-1) mean
value of the two colorless spinors is significantly reduced, and possibly proportional
to $\exp{\{-M_N|y_0-x_0|\}}$, then a good statistical precision is reached with a
number of updates of the lattice $(n_{0} \cdot n_{1})\propto \exp{\{(M_N-3 M_\pi/2)|y_0-x_0|\}}$. Notice
again that the factor at the exponent is halved with respect to the standard
Monte Carlo procedure. The remaining correction can be computed by a two-level algorithm
with a succession of simulations with larger and larger $\bar\Gamma_0$.
Also in this case the real effectiveness of the multi-level can only be
quantified by a realistic numerical test, see below.

\section{Numerical tests for the disconnected pseudoscalar propagator\label{sec:etapN}}
We test the ideas discussed in section \ref{sec:etap} in the quenched
approximation of QCD. We discretize gluons and fermions with the Wilson action,
and we impose open and periodic boundary conditions in the time and spatial
directions respectively~\cite{Luscher:2011kk,Luscher:2012av}. The inverse coupling
constant is fixed to $\beta=6/g_0^2=6.0$, the length of each spatial direction
to $L=24$, and the time extent to $T=64$. The lattice spacing is $a=0.093$~fm as 
fixed by assuming a physical value of $0.5$~fm for the Sommer scale
$r_0/a=5.368$~\cite{Guagnelli:1998ud}. The up and down quarks are
degenerate. Their masses are fixed by the hopping parameter value $k=0.1560$,
corresponding to a pion of approximatively $455$~MeV~\cite{Allton:1996yv}.

Numerical simulations have been carried out with a modified version
of the openQCD code version 1.4 \cite{openQCD1.4,Luscher:2012av}. We have generated
$n_0=200$ level-0 independent gauge field configurations spaced by $400$ molecular-dynamics
units (MDUs) with the Hybrid Monte Carlo (HMC).
Following section \ref{sec:etap}, the lattice has been split at
$x_0^\text{cut}=32$ in two domains of equal size $\Gamma_0$ and $\Gamma^*_0$. For all level-0
background gauge fields, $n_1=100$ level-1 configurations were generated by updating the
two regions independently while keeping fixed the spatial links at $x_0^\text{cut}=32$. Also
for these updates we used the HMC by skipping $400$ MDUs between measurements, a very conservative
choice for which the generation of the level-1 configurations is still cheaper than
the computation of the Wick contractions. Within this setup, the correlator
in Eq.~(\ref{eq:sing}) is naturally decomposed as
\begin{equation}\label{eq:discnum}
  C_{P_d}(y_0,x_0) = C^{(\text{f})}_{P_d}(y_0,x_0) + C^{(\text{r}_1)}_{P_d}(y_0,x_0) + C^{(\text{r}_2)}_{P_d}(y_0,x_0)\; . 
\end{equation}
The fully factorized contribution can be written as 
\begin{equation}\label{eq:factsing}
  C^{(\text{f})}_{P_d}(y_0,x_0) = \frac{1}{L^3}
    \left\langle \left[ \sum_{\vec{x}} \tr\left\{ \gamma_5 D^{-1}_{\Gamma_0}(x,x) \right\} \right]
                 \left[ \sum_{\vec{y}} \tr\left\{ \gamma_5 D^{-1}_{\Gamma^*_0}(y,y)   \right\} \right] \right\rangle \;.
\end{equation}
The other two terms are given by 
\begin{align}
C^{(\text{r}_1)}_{P_d}(y_0,x_0) & =  \frac{1}{L^3} \sum_{\vec{x},\vec{y}} \left\langle W^{(\text{r}_1)}_{P_d}(y,x) +
(\Gamma_0,x) \leftrightarrow (\Gamma^*_0,y) \right\rangle\;,\\[0.25cm]
C^{(\text{r}_2)}_{P_d}(y_0,x_0) & =  \frac{1}{L^3} \sum_{\vec{x},\vec{y}} \left\langle W^{(\text{r}_2)}_{P_d}(y,x)
  \right\rangle\;,\label{eq:factsing3}
\end{align}
where $W^{(\text{r}_1)}_{P_d}(y,x)$ and $W^{(\text{r}_2)}_{P_d}(y,x)$ are defined in Eqs.~(\ref{eq:rall0}) and
(\ref{eq:rall}). If the Wilson--Dirac
operator is written as
\begin{equation}
  2\kappa D = \id - \kappa D_\text{hop} \;,
\end{equation}
the trace can be re-expressed as\footnote{For the $O(a)$-improved Wilson-Dirac operator
$p\leq 2$.}~\cite{Thron:1997iy,McNeile:2000xx,Bali:2009hu}.
\begin{equation}\label{eq:HPEtr}
  \tr\left\{\gamma_5 D^{-1}\right\} = \kappa^p \tr\left\{\gamma_5 D_\text{hop}^p D^{-1}\right\}  \qquad p \leq 8 \;.
\end{equation}
By choosing $p=8$, all the traces appearing in the contributions (\ref{eq:factsing})--(\ref{eq:factsing3})
have been estimated stochastically, e.g.\
\begin{equation}
  \sum_{\vec{x}} \tr\left\{\gamma_5 D^{-1}(x,x)\right\} \longrightarrow \frac{1}{n_\text{src}}\sum_{i=1}^{n_\text{src}} \sum_{\vec{x}} \eta_i^\dagger(x) \left[\kappa^8 D_\text{hop}^8 D^{-1} \gamma_5 \eta_i\right](x) \;,
\end{equation}
by inverting the various Dirac operators on the very same $n_\text{src}=100$
Gaussian random sources $\eta_i$ \cite{Sommer:1994gg,Foster:1998vw}, defined on the whole
space-time volume\footnote{For the factorized contribution the $\eta_i$
acts effectively as two independent random sources, one for each domain. The estimate of the two traces is thus
obtained with a single global inversion per random source.},
and by contracting the solution with a time-slice of $\eta_i$. The hopping parameter expansion,
used in Eq.~(\ref{eq:HPEtr}), reduces the variance of the stochastic estimator significantly. Other
techniques~\cite{DeGrand:2004qw,Giusti:2004yp,Foley:2005ac,Blum:2012uh}
may further reduce the cost of the computation, but we prefer to keep it simple and 
focus on factorization.
\begin{figure}[p]
  \centerline{\includegraphics{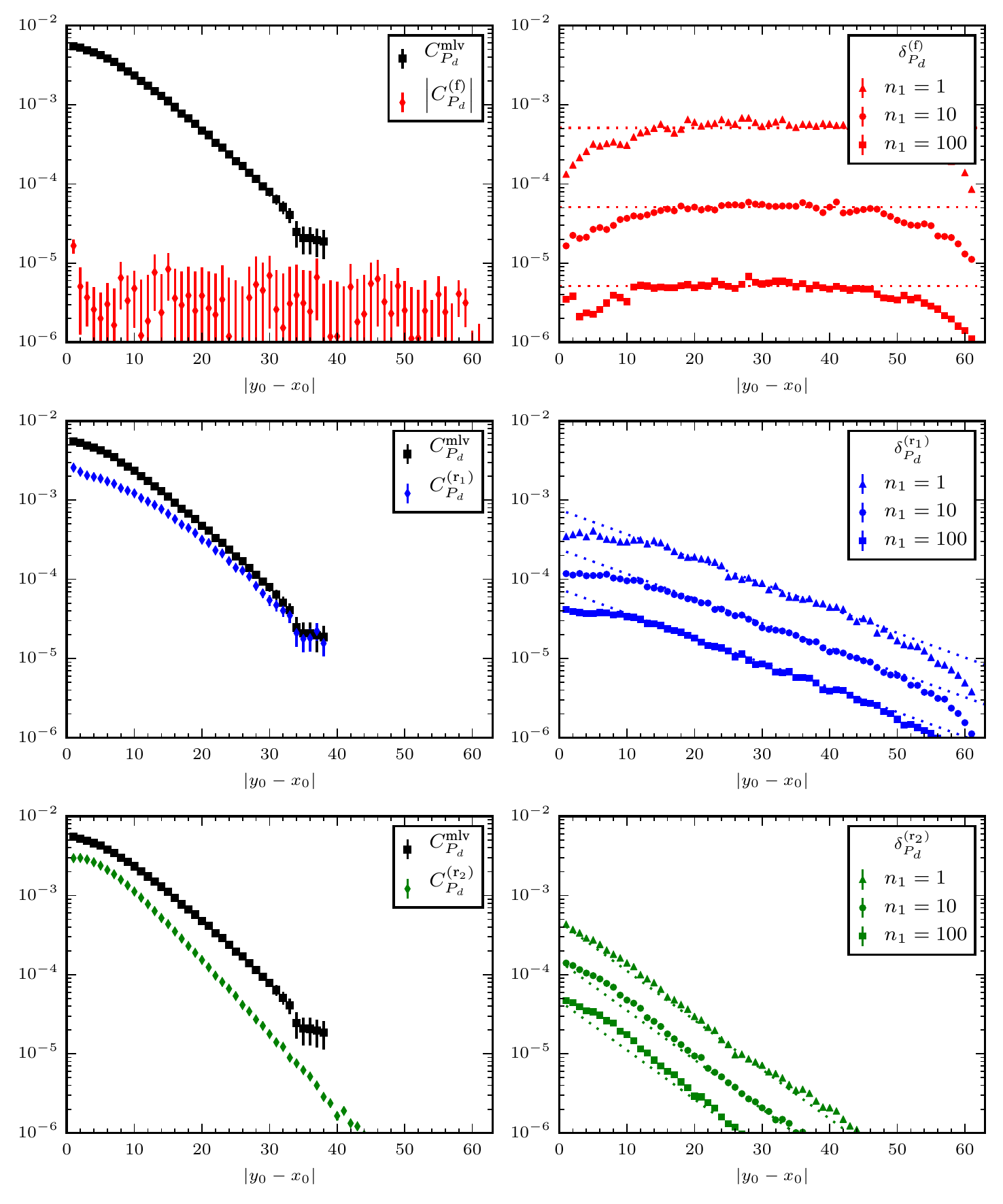}}
\caption{Left-column plots: the three contributions on the r.h.s.\ of
Eq.~(\protect\ref{eq:discnum}) are shown,
together with the best estimate of the full correlator (the sum of the three),
as a function of the time separation $|y_0-x_0|$.
Right-column plots: the errors of the various contributions are shown
as a function of the time distance for various values of $n_1$.}
\label{fig:singlet1}
\end{figure}

The $C^{(\text{f})}_{P_d}$ contribution is estimated by first averaging,
for each of the level-0 configurations, the two traces independently over the $n_1$
level-1 background fields. The expectation value of the product of the two means
is then computed by averaging over the $n_0$ configurations.
The other two contributions are computed as if
the $n_1$ (subset of) configurations, generated for each of the level-0 boundary fields,
were correlated level-0 ones. The  measures  of $C^{(\text{r}_1)}_{P_d}$ and
$C^{(\text{r}_2)}_{P_d}$ are thus grouped in bins of $n_1$, and the expectation values
and their errors are determined as usual by treating the bins as $n_0$ independent
measurements.

\subsection{Numerical results}
The numerical results for $C^{(\text{f})}_{P_d}$, $C^{(\text{r}_1)}_{P_d}$, and
$C^{(\text{r}_2)}_{P_d}$ are plotted in
Fig.~\ref{fig:singlet1} as a function of the time separation of the
pseudoscalar densities. The central values and their errors are shown in the plots
on the left and right columns respectively. The best estimate of $C_{P_d}$
(the sum of the three) is also shown in each plot on the left
for comparison. In all cases  $x_0$ and $y_0$ belong to different domains, $y_0>x_0$, and they
are chosen to be as much as possible equidistant from $x_0^\text{cut}$.

The statistical error on $C_{P_d}^{(\text{f})}$, top-right plot of Fig.~\ref{fig:singlet1}, is
a flat function of $|y_0-x_0|$ with sizeable deviations near the boundaries of the
domains. Error bars are smaller than the symbols. Up to
the largest value that we have, $n_1=100$, the error decreases as $n^{-1}_1$, i.e.\ the two-level
Monte Carlo works at full potentiality. The mean value of $C_{P_d}^{(\text{f})}$, top-left plot,
is compatible with zero. The correlation between
$C_{P_d}$ and $C_{P_d}^{(\text{f})}$ goes from $0.9$ to $1.0$ when $|y_0-x_0|$ varies from $15$ to
$50$, a value which collapses toward zero when the multi-level is switched on.
\begin{figure}[t!]
\centerline{\includegraphics{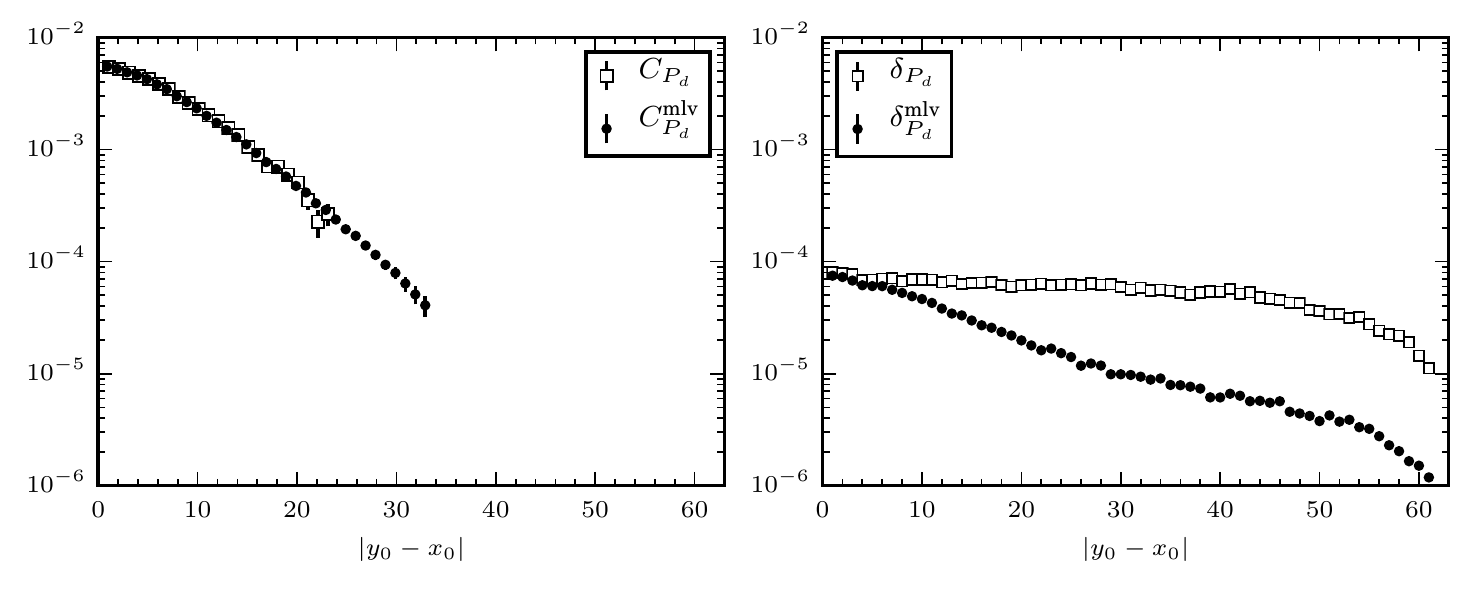}}
\caption{The best estimate of $C_{P_d}(y_0,x_0)$ (left) and of its error (right)
are shown as a function of the time distance, with and without two-level integration of
the factorized contribution. In the latter case the $n_1$ (subset of) configurations,
generated for each of the level-0 boundary fields, are treated as if
they were correlated level-0 ones. The $n_1$ measures are thus binned together,  
and the mean and its error are computed as usual by treating the bins as independent.}
\label{fig:singlet2}
\end{figure}

The statistical error on $C_{P_d}^{(\text{r}_1)}$, middle-right plot of Fig.~\ref{fig:singlet1},
shows a strong dependence on $|y_0-x_0|$. It is compatible with an exponential behavior of the form
$\exp\{-M |y_0-x_0|/2\}$ with an effective mass $M=0.14$, i.e.\ lighter than expected and
roughly $2/3$ of the pion mass\footnote{We did not
attempt to study the dependence of this parameter on the finite size or other sources of systematics.}.
It decreases as $n^{-1/2}_1$ up to $n_1=100$ and, at fixed time distance, it becomes the dominant contribution to
the error of $C_{P_d}$ once a large enough number $n_1$ of level-1 updates have been carried out.
The mean value of $C_{P_d}^{(\text{r}_1)}$ is roughly
$2/3$ of the full correlator at $|y_0-x_0|=15$, and it becomes
the dominant contribution up to $|y_0-x_0|=33$, after which the signal is lost. The statistical errors
of $C_{P_d}^{(\text{r}_2)}$ decreases exponentially as $\exp\{-M |y_0-x_0|\}$, and it scales
as $n^{-1/2}_1$.

A clear picture emerges from the above analysis. At large time distances,
the statistical error on the standard estimate of the disconnected pseudoscalar
propagator is dominated by the one on $C_{P_d}^{(\text{f})}$. The second largest
contribution is the statistical error on $C_{P_d}^{(\text{r}_1)}$ which, however,
is exponentially suppressed as $\exp\{-M |y_0-x_0|/2\}$. Once the two-level integration
is switched on, the error on $C_{P_d}^{(\text{f})}$ decreases as $n_1^{-1}$, while the one on
$C_{P_d}^{(\text{r}_1)}$ continues to scale as\footnote{A two-level algorithm can be
used to further reduce the statistical error on $C_{P_d}^{(\text{r}_1)}$ by
a domain decomposition of the exact inverse $D^{-1}$ in Eq.~(\ref{eq:rall0})
with the cut at, for instance, $x_0=40$. This is an improvement which
goes beyond the exploratory numerical study of this work.}
$n_1^{-1/2}$. The parameter $n_1$ can thus
be tuned, up to a prefactor of $\mathcal{O}(1)$, so that
$n_1 \sim \exp\left\{ M  d \right\}$ with 
$d$ being the maximum temporal distance in which one is
interested in\footnote{The contributions $C_{P_d}^{(\text{f})}$ and $C_{P_d}^{(\text{r}_1)}$ can be computed
with different number of sources, different value of $p$ for the HPE, etc. The prefactor of
this estimate can thus change depending on the details of the computation.}. This way the error on the factorized contribution is reduced to the level of (or below) the uncertainty on
$C_{P_d}^{(\text{r}_1)}$ at the same cost of generating $n_0\cdot n_1$ global configurations.
The net computational gain is therefore $\sim n_1$, and a good statistical precision is
reached with a number of updates  $(n_{0} \cdot n_{1})\propto \exp{\{M_\pi|y_0-x_0|\}}$.
Notice that the factor at the exponent is halved with respect to the standard Monte Carlo.

Our best estimate of the disconnected pseudoscalar propagator
is shown in Fig.~\ref{fig:singlet2}, where also the result without the
multi-level is reported for comparison. Using the two-level algorithm,
the signal-to-noise ratio remains larger than 1 for ten additional time slices.
This is better seen in the right plot, where
the statistical error is shown in the two cases. With the standard Monte Carlo
the error is approximatively flat, while for the two-level algorithm decreases
exponentially. The reduction is evident from distance $15$, and becomes
approximatively $n_1^{1/2}=10$ at the point $x_0=30$ that was taken to fix
$n_1$. For $n_1=100$, the overall gain in the computational cost is approximatively
50 since we have to invert two time the Wilson--Dirac operator on each random
source.

\section{Numerical tests for the pion propagator\label{eq:numtest}}
We have tested the factorized approximation of the quark propagator within the same
lattice setup of the previous section. The number of independent gauge
configurations have been increased to $1000$. For each of them $10$ Gaussian random
sources have been generated on the time-slice at
$x_0=4$, and the exact and the approximated quark propagators, as defined in
Eqs.~(\ref{eq:approxlong}) and (\ref{eq:approxlong2}), have been computed on the sources.
\begin{figure}[p]
  \centerline{\includegraphics{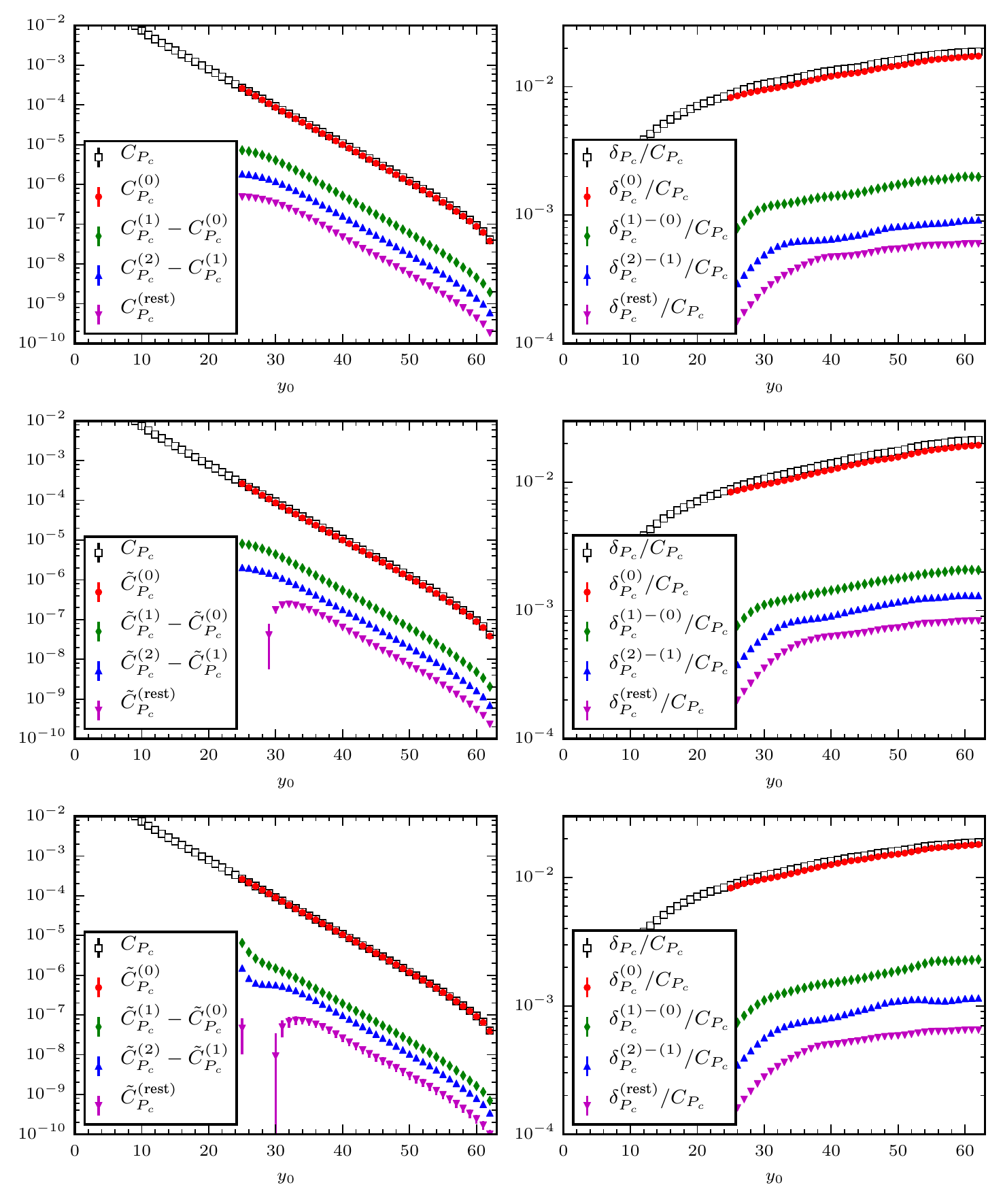}}
\caption{Top-line plots: central values (left) and their statistical errors (right) of the
five terms appearing on both sides of Eq.~(\protect\ref{eq:seriespi}). Middle-line plots:
analogous results but for a factorized approximation where a projector on the deflation
subspace has been inserted to cut the fermion lines. Bottom-line plots: the same 
but with fermion lines cut by a projector defined via $120$ modes computed by the
inverse iteration technique, see main text.}
\label{fig:pp_factorized}
\end{figure}
The corresponding pion propagators have then been calculated by contracting the
indices and averaging over the random sources as usual. The very same sources have
been used for $C_{P_c}(y_0,x_0)$ in
Eq.~(\ref{eq:pipi}), for
\be\label{eq:decpi}
C^{(X)}_{P_c}(y_0,x_0)  = 
- \frac{1}{L^3}\, \sum_{{\vec x}, {\vec y}}\, \langle W^{(X)}_{P_c}(y,x)\rangle \qquad
X=\text{f}, \text{r}\; ,\\
\ee
and for the analogous correlators $\tilde C^{(\text{f})}_{P_c}$ and $\tilde C^{(\text{r})}_{P_c}$.
Similarly to Eq.~(\ref{eq:Dseries}), the pion propagator can be expanded as 
\be\label{eq:seriespi}
C_{P_c} = C^{(0)}_{P_c} +
\Big[C^{(1)}_{P_c} - C^{(0)}_{P_c}\Big] +
\Big[C^{(2)}_{P_c} - C^{(1)}_{P_c}\Big] +
C^{(\text{rest})}_{P_c}\; .   
\ee
where $C^{(i)}_{P_c}$ for $i=0,1,2$ is a succession of factorized propagators $C^{(\text{f})}_{P_c}$
as defined in Eq.~(\ref{eq:decpi}) for $x_0^{\rm cut}=24$ and for $\Delta=8,12,16$ respectively,
while $C^{(\text{rest})}_{P_c} = C^{(\text{r})}_{P_c}$ for $\Delta=16$.

The correlation between $C_{P_c}(y_0,4)$ and $C^{(\text{f})}_{P_c}(y_0,4)$ turns out to be practically
1 for all $y_0$ and for all three values of $\Delta$. In the plots on the top of
Fig.~\ref{fig:pp_factorized}
we show the central values and the statistical errors of the five terms appearing on
both sides of Eq.~(\ref{eq:seriespi}) as a function of $y_0$. The statistical errors
are normalized to $C_{P_c}(y_0,4)$. If we pick up a typical point, $y_0=40$, we get
\begin{align}
\label{eq:expdelta}
C^{(0)}_{P_c} & =  (101.1 \pm 1.3)\cdot 10^{-7}\; , \nonumber\\[0.25cm]
\Big[C^{(1)}_{P_c} - C^{(0)}_{P_c}\Big] & = 
(5.22 \pm 0.15)\cdot 10^{-7}\; , \nonumber\\[0.25cm]
\Big[C^{(2)}_{P_c} - C^{(1)}_{P_c} \Big]& = 
(1.59 \pm 0.07)\cdot 10^{-7}\; ,\\[0.25cm]
C^{(\text{rest})}_{P_c} & =  (0.48\pm0.05)\cdot 10^{-7}\; .\nonumber
\end{align}
All these results show that, for this quark mass, the factorized correlator
approximates the exact one at the level of $5\%$ already for $\Delta=8$, a precision
which increases by one order of magnitude for $\Delta=16$. The reduction of the central
value of $[C^{(i)}_{P_c} - C^{(i-1)}_{P_c}]$ is in line
(even a bit faster) with the expectations from Eq.~(\ref{eq:errorfact}),
while the decreasing of its statistical error is a bit slower. 

The two plots in the middle of Fig.~\ref{fig:pp_factorized} show analogous
results but with the factorized propagators computed by inserting $\tilde S^{(\text{f})}(y,x)$,
as defined in Eq.~(\ref{eq:approxlong2}), in the contraction. The set of orthonormal
vectors $\phi_i$ are chosen to be those which form the deflation subspace generated
from $N_s=60$ global modes as defined in Ref.~\cite{Luscher:2007se}. The cut is again at
$x_0^{\rm cut}=24$, and $\Delta=8,12,16$. It is clear that the contribution from the deflation subspace
saturates nicely the exact pion correlator, and that the factorization combines well with
deflation provided the number of modes is large enough.

Finally in the plots on the bottom of Fig.~\ref{fig:pp_factorized} we
show analogous data but with the $\phi_i$ being $N_m=120$ orthonormal vectors
computed by applying $10$ inverse iterations of the block Wilson--Dirac operator,
defined in the thick time-slice $\Omega^*_{i_c-1}=\Lambda_{i_c-1}\cup\Lambda_{i_c}$, on
randomly chosen vectors. In this case the leading contribution
$\tilde C^{(\text{f})}_{P_c}$ can be written as a sum of a manageable number of 
terms defined as in Eqs.~(\ref{eq:fullfactpi})--(\ref{eq:Bpions}). Since the $\phi_i$
depend on the gauge links belonging to $\Omega^*_{i_c-1}$ only, each term in the sum
could have been computed with a two-level Monte Carlo by keeping frozen the links in
$\Omega^*_{i_c-1}$ during the
level-1 updates. It is our experience, however, that for the pion propagator this
is not cost effective as for baryons, see below. It must also be said that, while
$N_m=120$ is good enough for the lattice setup we have chosen, this number
may have to be increased significantly at larger volumes. 

\section{Tests of two-level integration for the baryon propagator}
We have computed the baryon propagator on the same $n_0=1000$ configurations
used for the pion in section~\ref{eq:numtest}. Also in this case we have
calculated the exact correlator $C_N(y_0,x_0)$, and the various
contributions defined as
\be\label{eq:baryN}
C^{(X)}_N(y_0,x_0) = \sum_{{\bf y}}
\Big\{\langle W^{(X)}_{N1}(y,x) \rangle - \langle W^{(X)}_{N2}(y,x)\rangle \Big\}\; ,
\qquad X=\text{f}, \text{r}\; 
\ee
by using the exact quark propagator and the factorized approximations as defined in
Eqs.~(\ref{eq:approxlong}) and (\ref{eq:approxlong2}). All of them have been determined
starting from 4 local sources located at a randomly chosen spatial position
on the time-slice at $x_0=4$.
\begin{figure}[p]
\centerline{\includegraphics{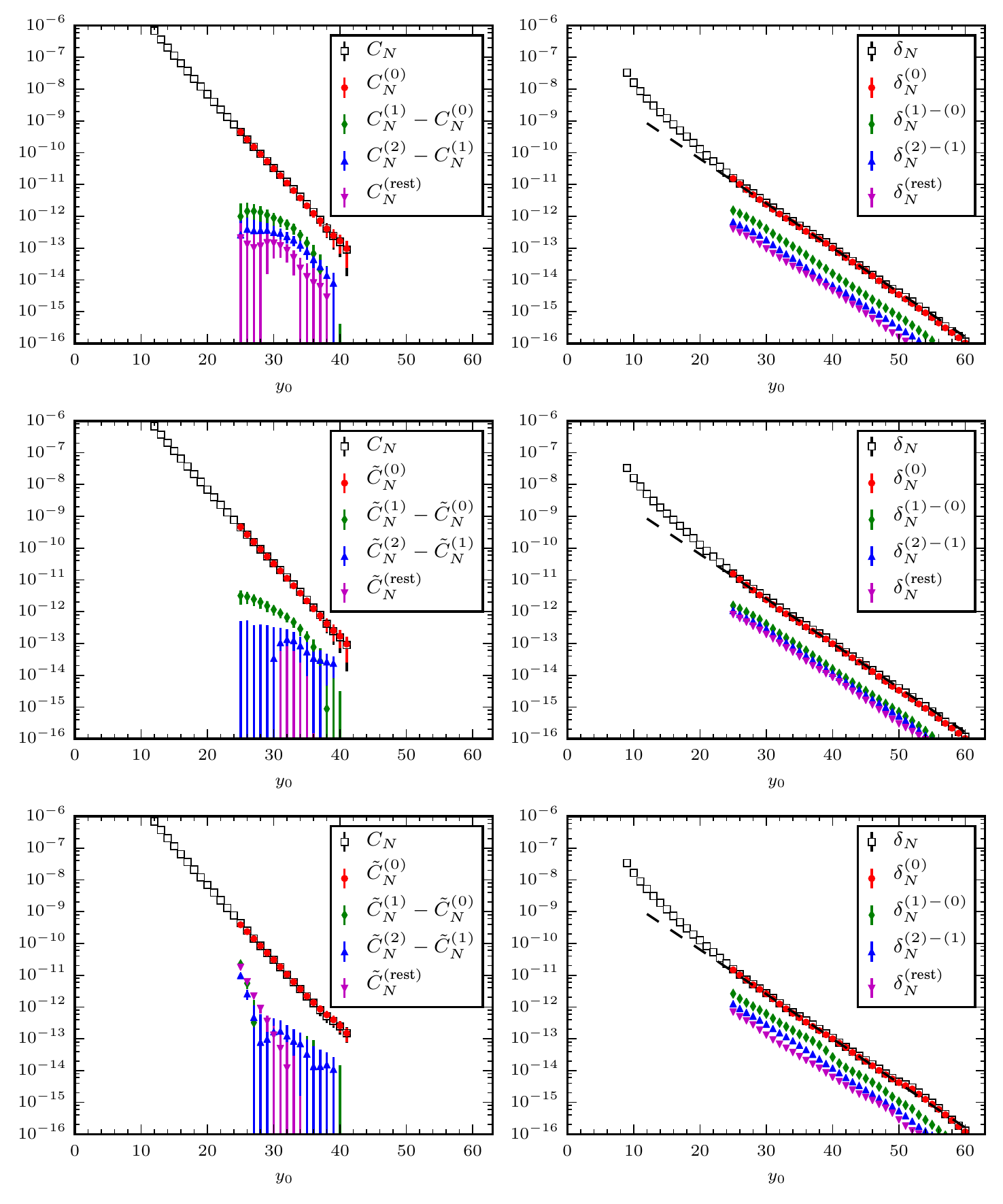}}
\caption{Top-line plots: central values (left) and their statistical errors (right) of the
  five terms appearing on both sides of Eq.~(\protect\ref{eq:seriespi}). For clarity in
data are shown only up to $x_0=41$, after which the signal for the correlator is lost.
Middle-line plots: analogous results but for a factorized approximation where a projector on the deflation
subspace has been inserted to cut the fermion lines. Bottom-line plots: the same 
but with fermion lines cut by a projector defined via $120$ modes computed by the
inverse iteration technique, see main text.
}
\label{fig:nn_factorized}
\end{figure}

As for the pion, we expand the nucleon propagator as 
\be\label{eq:seriesN}
C_{N} = C^{(0)}_{N} +
\Big[C^{(1)}_{N} - C^{(0)}_{N}\Big] +
\Big[C^{(2)}_{N} - C^{(1)}_{N}\Big] +
C^{(\text{rest})}_{N}\; .   
\ee
where $C^{(i)}_{N}$ for $i=0,1,2$ is a succession of factorized correlators $C_{N}^{(\text{f})}$
as defined in Eq.~(\ref{eq:baryN}) for $x_0^{\rm cut}=24$ and for $\Delta=8,12,16$ respectively,
while $C^{(\text{rest})}_{N} = C^{(\text{r})}_{N}$ for $\Delta=16$. The correlation between $C_{N}(y_0,4)$
and $C^{(\text{f})}_{N}(y_0,4)$ is practically 1 for all $y_0$ and for all three values of $\Delta$
also for the nucleon.
\begin{figure}[t!]
\centerline{\includegraphics{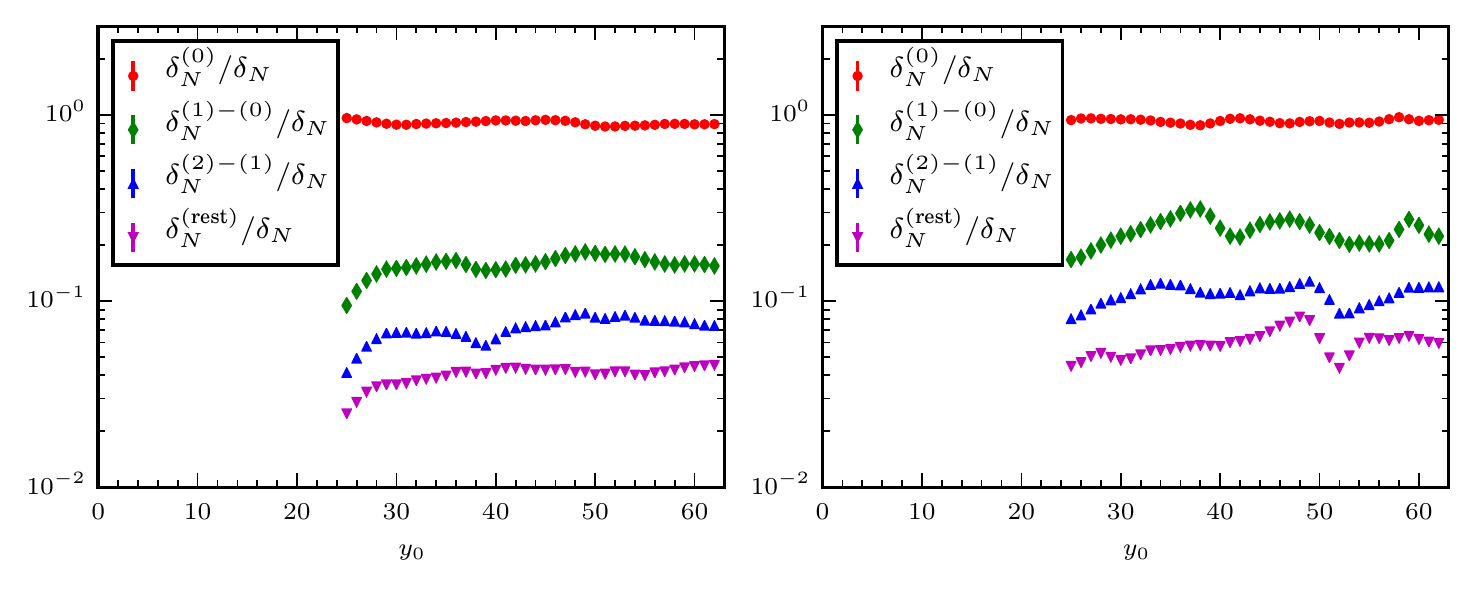}}
\caption{Left: statistical errors of the
four terms appearing on the r.h.s.\ side of Eq.~(\protect\ref{eq:seriespi})
normalized to the error of the exact propagator on the l.h.s.\ of the
same equation. Right: analogous results but for a factorized
approximation where the fermion lines are cut by the projector defined
via the $120$ modes computed by the inverse iteration technique, see main text.
}
\label{fig:nn_fact_diferr}
\end{figure}
In the plots on
the top of Fig.~\ref{fig:nn_factorized} we show the central values (left) and the statistical
errors (right) of the five terms appearing on both sides of Eq.~(\ref{eq:seriesN})
as a function of $y_0$. Due to the exponential suppression of the signal with
respect to the noise the data on the left plot has large statistical errors, especially the
smallest three contributions. On the right plot the statistical error of the
exact correlator (and of all the others) follows the expected exponential behavior
$\propto e^{-3 M_\pi|y_0-4|/2}$ (black line) with $M_\pi=0.215$ from $y_0\gtrsim 25$,
fully confirming the analysis in Ref.~\cite{Parisi:1983ae}. The hierarchy among the statistical
errors of the various terms is evident in the
left plot of Fig.~\ref{fig:nn_fact_diferr}, where the errors are normalized
to the one of the exact correlator. If we pick up a typical point,
$y_0=35$ not to lose the signal for the correlator, we get
\begin{align}
\label{eq:expdeltaN}
C^{(0)}_{N} & =  (21.3 \pm 4.6)\cdot 10^{-13}\; , \nonumber\\[0.25cm]
\Big[C^{(1)}_{N} - C_N^{(0)}\Big] & = 
(1.44 \pm 0.83)\cdot 10^{-13}\; , \nonumber\\[0.25cm]
\Big[C^{(2)}_{N} - C^{(1)}_{N} \Big]& = 
(0.78 \pm 0.35)\cdot 10^{-13}\; ,\\[0.25cm]
C^{(\text{rest})}_{N} & =  (0.13 \pm 0.20)\cdot 10^{-13}\; .\nonumber
\end{align}
At this quark mass, the factorized correlator
approximates the exact one at the level of $5-10\%$ already for $\Delta=8$, a precision
which increases which $\Delta=16$ even though the statistical errors are too large to
justify a more precise statement. The reduction
of the statistical error from top to bottom in 
Eqs.~(\ref{eq:expdeltaN}) is more than a factor 20, and it seems to 
decrease a bit slower than the expectation from Eq.(\ref{eq:errorfact}). 

The other four plots in Fig.~\ref{fig:nn_factorized} are analogous to the ones for the pions.
The two plots in the middle show the results for the factorized
propagators computed by inserting in the contraction $\tilde S^{(\text{f})}(y,x)$ as defined
in Eq.~(\ref{eq:approxlong2}). The set of orthonormal vectors $\phi_i$ are chosen to be
those which form the deflation subspace. The cut is again at
$x_0^{\rm cut}=24$, and $\Delta=8,12,16$. The remarkable fact is that
the contribution from the deflation subspace saturates nicely the exact nucleon
correlator, provided the number of modes is large enough. The factorization combines
well with deflation. 

Finally in the plots on the bottom of Fig.~\ref{fig:nn_factorized} we
show analogous data but with the $\phi_i$ being $N_m=120$ orthonormal vectors
computed as for the pions. The statistical errors normalized to the one
of the full correlator are shown on the right plot of Fig.~\ref{fig:nn_fact_diferr}.

\subsection{Two-level Monte Carlo}
The colorless spinors $\mathds{B}$ in Eqs.~(\ref{eq:W0tbary}) depend
on the gauge field belonging either to $\bar\Gamma_0$ or $\Gamma^*_0$.
When the links in $\Omega^*_{i_c}$ are kept frozen, the dependence of the action
and of the approximated contractions on them is factorized. We can thus compute
$C^{(\text{f})}_N(y_0,x_0)$ with a two-level Monte Carlo. For $n_0=50$ level-0
configurations (the first $50$ of the $1000$ generated previously), we
generate $n_1=20$ level-1 gauge fields\footnote{When $n_1$ is not a large number,
it is feasible to avoid the cut of the fermion lines by computing the factorized
propagator on the $n_1^2$ combinations of level--1 configurations.} by freezing
the link variables in $\Omega^*_{i_c}$.
\begin{figure}[t!]
\centerline{\includegraphics{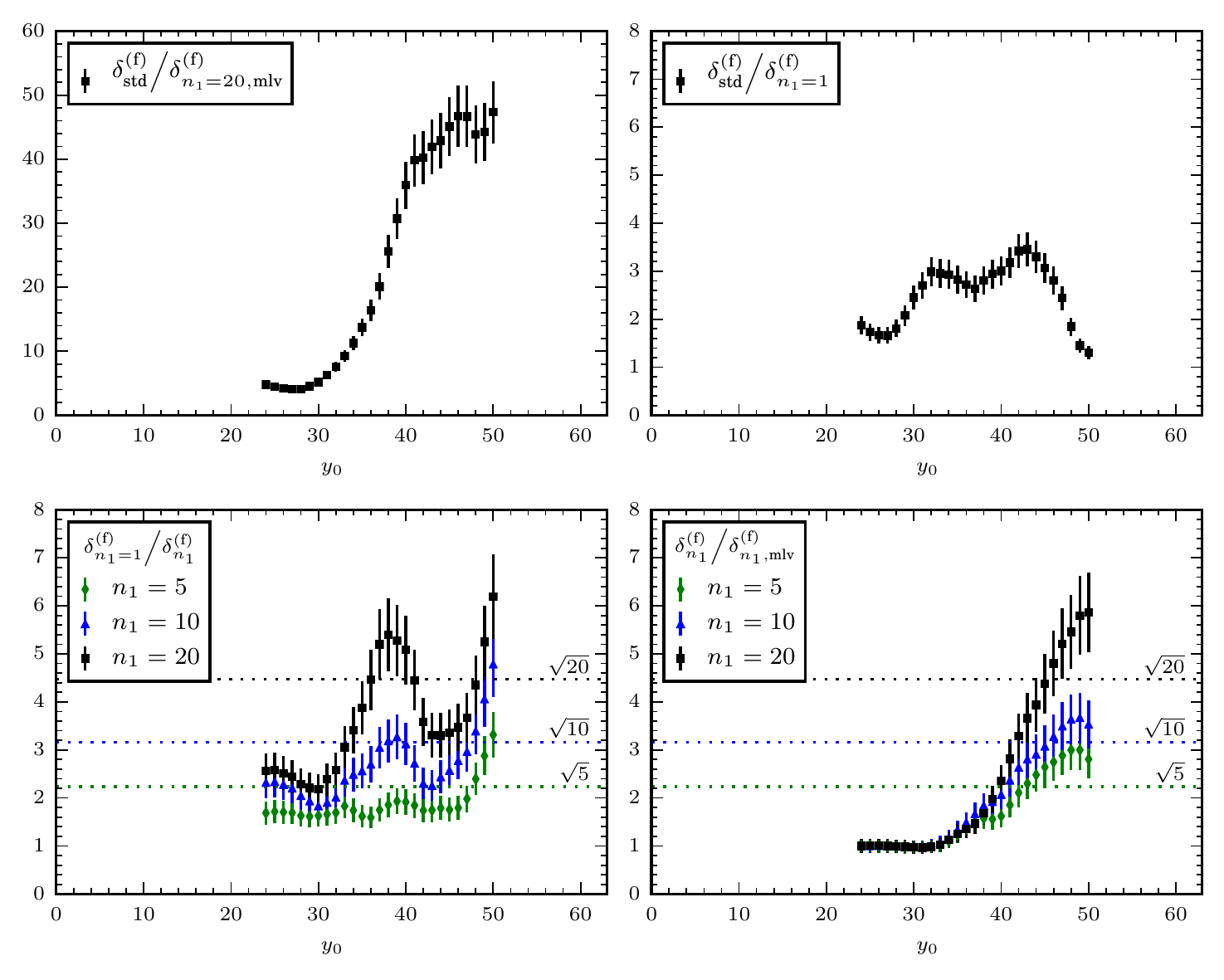}}
\caption{The total gain on the statistical error of
$C^{(\text{f})}_N(y_0,x_0)$ in the two-level Monte Carlo (top-left) is due to various
factor: the sum over all the points on the time-slice at $x_0=4$ (top-right), the
averaging over the $n_1$ level-1 configurations (bottom-left), the two-level averaging
of the $\mathds{B}$-spinors (bottom-right). See main text for more details.}
\label{fig:gain}
\end{figure}
The thick time-slice averages of the colorless spinors are then performed independently
on the level-1 configurations, and the matrix elements of
$\gamma_5 P_-$ between the spinors is averaged over the 50 level-0 boundary fields.

In the top-left plot of Fig.~\ref{fig:gain} we show the total reduction achieved
for the statistical error on $C^{(\text{f})}_N(y_0,x_0)$. In particular what is plotted in this graph is
the ratio of (a) the statistical error on the factorized correlator -- computed
by averaging over the $1000$ level-0 configurations at our disposal and the 4 local
sources and finally multiplied by $\sqrt{1000/50}$ -- over (b)
the statistical error achieved with the two-level integration. For $y_0\gtrsim 30$ we observe
a sharp reduction of the error by a factor which becomes $40$--$50$ for $y_0\gtrsim 40$.
The cost per level-0 configuration,
without counting scalar products and updates, is roughly $70$ times higher.
For a given target statistical error, this results in a net reduction of the cost of the
simulation of $20$--$40$ times.

The origin of the gain is due to various factors: sum over all the points on the time-slice
at $x_0=4$, averaging over the $n_1$ level-1 configurations, two-level averaging. On the
top-right plot we show the ratio of the errors of the
local estimator, averaged over the 4 local sources and multiplied by $\sqrt{1000/50}$, over
the one on $C^{(0)}_N(y_0,x_0)$ obtained with the $50$ level-0 configurations only. After a
few transient time-slices, the gain is in the range $2$--$3$. On the bottom-left
plot we show the gain due to the averaging over the $n_1$ gauge configurations but without
two-level integration. We just bin the $n_1$ measurements for each level-0
configurations, and we compute the errors considering the bins as independent. After few 
time-slices were no gain is observed due to the freezing of the links belonging to
$\Omega^*_{i_c}$, the gain scales approximatively as $\sqrt{n_1}$. In the bottom-right plot the
reduction of the error due to the two-level independent averaging is shown.
For $y_0\gtrsim 35$ the gain is clearly visible, and at larger distances saturates
the expected $\sqrt{n_1}$ factor up to  $n_1=20$.
\begin{figure}[t!]
\centerline{\includegraphics{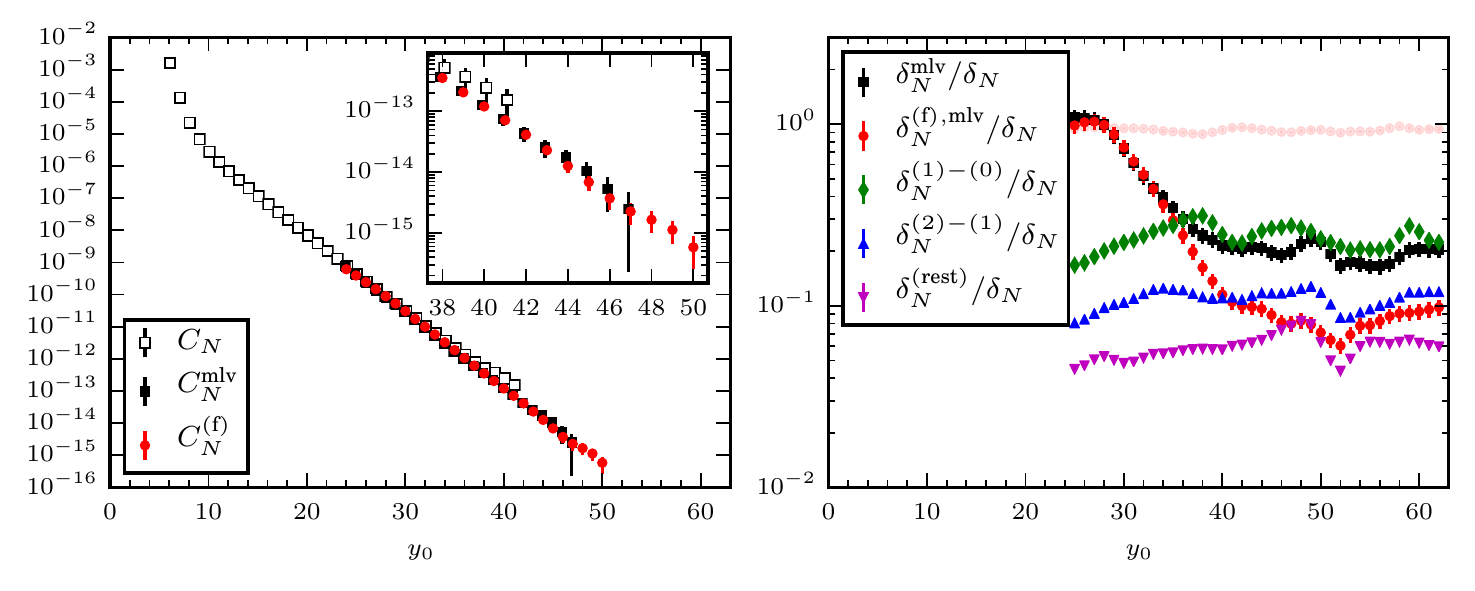}}
\caption{Left: best results for $C_N(y_0,x_0)$ with and without two-level integration,
and for $C^{(0)}_N(y_0,x_0)$. Right: same as in the right plot of
Fig.~\protect\ref{fig:nn_fact_diferr} but with the error on $C^{(0)}_N(y_0,x_0)$
from the two-level Monte Carlo.  For completeness we show also the statistical error on our best
two-level estimate of the exact correlator}
\label{fig:final}
\end{figure}

The final results for the correlator with and without the two-level integration are shown
in the left plot of Fig.~\ref{fig:final}. In the two-level Monte Carlo 
the signal-to-noise ratio for the factorized contribution remains larger than
one for 10 additional time-slices with respect to the standard evaluation. When we add
the rest of the correlator, $C^{(\text{r})}_{N}$, the gain reduces to 5 additional points.
The effectiveness of the two-level integration is better seen on the right plot
of Fig.~\ref{fig:final}, which is a replica of the right graph in Fig.~\ref{fig:nn_fact_diferr}
but with the error on $C^{(0)}_N$ coming from the two-level integration
(error with standard Monte Carlo also shown with shadow red points). For completeness
we report also the statistical error on our best two-level estimate of the exact correlator,
normalized to the one obtained with the standard Monte Carlo.

A rather clear picture emerges, similar to what we discussed for the disconnected correlator.
At large time distances, the statistical error on the standard estimate of $C_N$ is dominated
by the one on $C^{(0)}_N$. Once the two-level integration
is switched on, the error on $C^{(0)}_N$ decreases roughly as $n_1^{-1}$, while the one on
the rest of the correlator continues to scale as $n_1^{-1/2}$. We thus tuned $n_1$ so that
the error on $C^{(0)}_N$ is smaller (roughly $1/2$) of the uncertainty on $\Big[C^{(1)}_{N} - C_N^{(0)}\Big]$
at the same cost of generating $n_0\cdot n_1$ global configurations. If one wants to
gain further (beyond the scope of this paper),
the leading correction $\Big[C^{(1)}_{N} - C^{(0)}_{N}\Big]$ needs also to be integrated
with a two-level algorithm with $\Delta=12$.  This way the error on 
this contribution can be reduced at the level or below the one on $C^{(0)}_N(y_0,x_0)$
and $\Big[C^{(2)}_{N} - C^{(1)}_{N}\Big]$ , and so on. A good statistical precision can thus
be reached with a number of updates of the lattice
$(n_{0} \cdot n_{1})\propto \exp{\{(M_N-3 M_\pi/2)|y_0-4|\}}$, i.e.\ a factor at the exponent
which is halved with respect to the one in the standard Monte Carlo procedure.

\section{Conclusions}
The numerical computation of many interesting hadronic correlation functions in lattice
QCD suffers from signal-to-noise ratios which decrease exponentially with the time
distance of the sources. Notable examples are meson correlators with disconnected contributions,
static-light correlators, baryonic correlation functions with and without
disconnected Wick contractions, etc. Based on the experience in purely bosonic
theories, our physics intuition would suggest that multi-level algorithms
would lead to an impressive acceleration of those computations, opening new
perspectives in lattice QCD.

Formulating multi-level integration schemes in systems
with fermions, however, is not as straight forward as for bosons. The gauge-field
dependence of the fermion determinant and of the propagator need to be
judiciously factorized before integrating the Wick contractions. Here we have shown
that this can be achieved in (quenched) QCD by properly combining the ideas
of multi-level integration and domain decomposition of the quark propagator.

The numerical tests that we have carried out for the disconnected correlator of two
flavor-diagonal pseudoscalar densities and for a nucleon two-point function
show indeed that the signal-to-noise ratio increases exponentially with the time
distance of the sources when a two-level integration is at work. In the very simple setup
that we have implemented, the number of configurations needed to reach a given statistical
precision is proportionally to the square root of those required in the standard case.

For the strategy to be useful in full QCD, the next step is to devise a similar
domain decomposition for the quark determinant. 

\section{Acknowledgments}
We thank M.~L\"uscher for several inspiring discussions. Many thanks to
Miguel Garcia Vera for discussions on various aspects of multi-level
integration. We thank the Galileo Galilei Institute for Theoretical Physics
for hospitality and the INFN for partial support during the beginning of this
work. Simulations have been performed on the PC-cluster PAX at DESY, Turing
and Wilson at Milano-Bicocca, and on Galileo at CINECA. We thank these institutions
for the computer resources and the technical support. L.G.~acknowledge partial
support by the INFN SUMA project. 

\appendix

\section{Wilson--Dirac operator\label{app:Dw}}
The massive Wilson--Dirac operator is defined as~\cite{Wilson:1975id}
\be\label{eq:wdo}
D = D_{\rm w} + m_0\; ,
\ee
where $m_0$ is the bare quark mass, $D_{\rm w}$ 
is the massless operator  
\begin{equation}
D_{\rm w}=
\frac{1}{2}\left\{ \gamma_\mu(\nabla^*_\mu+\nabla_\mu)- \nabla^*_\mu\nabla_\mu\right\}\; ,
\end{equation}
$\gamma_\mu$ are the Dirac matrices, and the
summation over repeated indices is understood. 
The covariant forward and backward derivatives 
$\nabla_\mu$ and $\nabla^*_\mu$ are defined to be
\be
\nabla_{\mu}\psi(x) =  U_\mu(x)\psi(x+\hat{\mu})-\psi(x)\; ,\quad
\nabla^*_\mu\psi(x) =  \psi(x)-  U_\mu^\dagger(x-\hat{\mu})
\psi(x-\hat{\mu})\; ,
\ee
where $U_\mu(x)$ are the link variables. The Wilson--Dirac
operator satisfies the $\gamma_5$-hermiticity relation
\be\label{eq:g5h}
D = \gamma_5 D^\dagger \gamma_5\; . 
\ee 

\section{LU factorization of the block-banded Wilson-Dirac operator\label{eq:appLU}}
The $LU$ factorization for block banded matrices leads to the simple result for
the Wilson--Dirac operator~\cite{Quarteroni:2000}
\begin{align}
\hspace{-0.5cm}\begin{pmatrix}
D_{\Lambda_{0,0}} & D_{\Lambda_{0,1}}& 0 & \dots  \\
D_{\Lambda_{1,0}} & D_{\Lambda_{1,1}}& D_{\Lambda_{1,2}} & \dots \\
0       & D_{\Lambda_{2,1}}& D_{\Lambda_{2,2}} \\
\vdots       &       &  & \ddots \\
\end{pmatrix}
=
\begin{pmatrix}
1&B_0 &0 &  \dots  \\
0& 1  & B_1 &  \dots \\
0& 0  & 1&   \\
\vdots       &       &  & \ddots \\
\end{pmatrix}
\begin{pmatrix}
A_0& 0& 0  & \dots  \\
D_{\Lambda_{1,0}} & A_1& 0 & \dots \\
0   & D_{\Lambda_{2,1}} & A_2 \\
\vdots       &       &  & \ddots \\
\end{pmatrix}\, ,
\label{e:lu}
\end{align}
where the block matrices $D_{\Lambda_{i,j}}$ are defined
in Eqs.~(\ref{eq:fact1})--(\ref{eq:fact3}), while the 
$A_i$ and the $B_i$ are defined uniquely in terms of the
$D_{\Lambda_{i,j}}$ by the following recursion relations
\begin{align}
A_{T-1}&=D_{\Lambda_{T-1,T-1}}\; ,\nonumber\\[0.25cm]
A_i&=D_{\Lambda_{i,i}}-D_{\Lambda_{i,i+1}} A_{i+1}^{-1} D_{\Lambda_{i+1,i}}\qquad\!\! (i=0,\dots,T-2)\; ,\label{e:B}\\[0.25cm]  
B_i&=D_{\Lambda_{i,i+1}} A_{i+1}^{-1} \qquad\qquad\qquad\qquad (i=0,\dots,T-2) \nonumber\; .
\end{align}
Using the factorization (\ref{e:lu}), the linear system $D\,\psi\,=\,\eta$
can be easily solved, again leading to recursion relations. Let us 
consider the case where the source $\eta$ is non-zero only on one thick 
time-slice $\Lambda_k$. Solutions for sources on multiple time slices can be
obtained by superposition. The system
\begin{align}
\begin{pmatrix}
1&B_0 &0 &  \dots  \\
0& 1  & B_1 &  \dots \\
0   & 0 & 1&   \\
\vdots       &       &  & \ddots \\
0&&&&1
\end{pmatrix}
\begin{pmatrix}
\chi_0\\
\chi_1\\
\vdots\\
\\
\chi_{T-1}
\end{pmatrix}
=
\begin{pmatrix}
0\\
\vdots\\
\eta_k\\
\vdots\\
0
\end{pmatrix}
\end{align}
is solved by 
\begin{align}
\label{eq:chi}
\chi_i=\begin{cases}
\displaystyle\left[\prod^{k-1}_{j=i}(-B_j)\right] \eta_k & \ i<k\\
\eta_k & i=k\\
0 & i>k\\
\end{cases}
\end{align}
where in the last line the obvious ordered product has to be taken.
Using Eq.~(\ref{e:B}) we can rewrite 
\begin{align}
\prod_{j=i}^{k-1}(-B_j) \,\eta_k=(-)^{k-i}\,(D_{\Lambda_{i,i+1}} A_{i+1}^{-1})\, \dots\, (D_{\Lambda_{k-1,k}} A_k^{-1})\,\eta_k\; ,
\end{align}
where for $i<k$ the $\chi_i$ have support only on the boundaries. By
solving the system
\begin{align}
\begin{pmatrix}
A_0& 0 & 0  & \dots  \\
D_{\Lambda_{1,0}}  & A_1& 0 & \dots \\
0   &  D_{\Lambda_{1,2}} & A_2&  \\
\vdots       &       &  & \ddots \\
\end{pmatrix}
\begin{pmatrix}
\psi_0\\
\vdots\\
\\
\psi_{T-1}
\end{pmatrix}
=
\begin{pmatrix}
\chi_0\\
\vdots\\
\\
\chi_{T-1}
\end{pmatrix}\; ,
\end{align}
we get the final result
\begin{align}
\psi_{0}&=A_{0}^{-1}\, \chi_{0}\; ,\nonumber\\
\psi_{i}&=A_i^{-1}(\chi_i-D_{\Lambda_{i,i-1}} \psi_{i-1}) \qquad (i=1,\dots,T-1) \; .
\label{eq:LUf}
\end{align}
As for the $\chi_i$, the second term in the parentheses on the r.h.s.\
of Eq.~(\ref{eq:LUf}) live on the boundaries. The matrix $A^{-1}_i$ 
propagates these two contributions into the center of the thick time-slice. The 
Eq.~(\ref{eq:LUf}) is the basis of the so-called Thomas algorithm 
for the solution of (block) banded linear systems~\cite{Quarteroni:2000}.

\subsection{The $2 \times 2$ case}
The previous derivation for the $2 \times 2$ block-banded Wilson-Dirac operator
\be
D = \left(\begin{array}{c@{~~}c@{~~}}
D_{\Lambda_{0,0}} & D_{\Lambda_{0,1}} \\[0.25cm]
D_{\Lambda_{1,0}} & D_{\Lambda_{1,1}}
\end{array}   \right) = 
\left(\begin{array}{c@{~~}c@{~~}}
I & \;\;D_{\Lambda_{0,1}} D_{\Lambda_{1,1}}^{-1} \\[0.25cm]
0  & I
\end{array}   \right)
\left(\begin{array}{c@{~~}c@{~~}}
S_{\Lambda_{0,0}} & 0 \\[0.25cm]
D_{\Lambda_{1,0}}  & D_{\Lambda_{1,1}} 
\end{array}   \right)
\; ,
\ee
where the Schur complement is defined as 
\be\label{eq:Schur}
S_{\Lambda_{0,0}} = D_{\Lambda_{0,0}} - D_{\Lambda_{0,1}} D_{\Lambda_{1,1}}^{-1} D_{\Lambda_{1,0}}\; , 
\ee
leads to
\be\label{eq:Scmpt2}
D^{-1} = 
\left(\begin{array}{c@{~~}c@{~~}}
S_{\Lambda_{0,0}}^{-1} &
-S^{-1}_{\Lambda_{0,0}} D_{\Lambda_{0,1}} D_{\Lambda_{1,1}}^{-1}
\\[0.25cm]
- D_{\Lambda_{1,1}}^{-1} D_{\Lambda_{1,0}} S^{-1}_{\Lambda_{0,0}} &
\;\; D_{\Lambda_{1,1}}^{-1} + D^{-1}_{\Lambda_{1,1}} D_{\Lambda_{1,0}} S^{-1}_{\Lambda_{0,0}} D_{\Lambda_{0,1}} D^{-1}_{\Lambda_{1,1}}
\end{array}   \right)\; .
\ee
It is worth noting that $S^{-1}_{\Lambda_{0,0}}$ is the exact block in the block 
inverse of $D$. By putting the Schur complement in the bottom-right block,
the analogous formula can be written as
\be\label{eq:Scmpt}
D^{-1} = 
\left(\begin{array}{c@{~~}c@{~~}}
D_{\Lambda_{0,0}}^{-1} + D^{-1}_{\Lambda_{0,0}} D_{\Lambda_{0,1}} S^{-1}_{\Lambda_{1,1}} D_{\Lambda_{1,0}} D^{-1}_{\Lambda_{0,0}}  & 
- D_{\Lambda_{0,0}}^{-1} D_{\Lambda_{0,1}} S^{-1}_{\Lambda_{1,1}}\\[0.25cm]
-  S^{-1}_{\Lambda_{1,1}} D_{\Lambda_{1,0}} D_{\Lambda_{0,0}}^{-1} & S^{-1}_{\Lambda_{1,1}}
\end{array}   \right)\; .
\ee
with $S_{\Lambda_{1,1}}$ defined as in Eq.~(\ref{eq:Schur}) but with $1 \leftrightarrow 0$.

\subsection{Approximate factorization}
The formulas (\ref{eq:chi}) and (\ref{eq:LUf}) factorize the solution vector in terms 
of thick time-slice matrix products. Only the matrices $A_i$ carry 
the dependence on the links that belong to several slices, via (nested) Schur complements. 
The Eq.~(\ref{eq:fctpr}) can
be derived from  Eqs.~(\ref{eq:chi}) and (\ref{eq:LUf}) by a systematic approximation
of the LU decomposition. As in section~\ref{sec:fact} we choose the source point
$x\in\Lambda_m$ and the sink $y\in\Lambda_l$ with $l>m$, see Fig.~\ref{Fig:fig3}. By using
Eq.~(\ref{eq:LUf}) it is easy to show that
\be\label{eq:psil}
\psi_l =(-1)^{m-l} A_l^{-1}D_{l,l-1}
\dots A_{m+1}^{-1}D_{m+1,m}\psi_m
\ee 
By approximating $A_i$ with $D_{\Omega^*_i}$ and $\psi_m$ with $D^{-1}_{\Omega_{m+2}}\eta_m$ we
arrive at Eq.~(\ref{eq:fctpr}).

\end{document}